\documentclass[
 reprint,
 twocolumn,
 aps,
 pra,
 preprintnumbers,
 superscriptaddress,
 floatfix
]{revtex4-2}

\usepackage[a4paper,margin=1in]{geometry}
\usepackage{amsmath,amssymb,amsthm,mathtools}
\usepackage{siunitx}
\usepackage{bm}
\usepackage{physics}
\usepackage{graphicx}
\usepackage{booktabs}
\usepackage{xcolor}
\usepackage{hyperref}
\usepackage{nicefrac}
\usepackage[nameinlink,capitalise]{cleveref}
\usepackage[T1]{fontenc}
\usepackage{inconsolata}

\newcommand{\myscale}{0.48}

\newcommand{\threej}[6]{\begin{pmatrix}#1&#2&#3\\#4&#5&#6\end{pmatrix}}
\newcommand{\braketBig}[3]{\big< #1 \big|#2\big|#3\big>}

\begin{document}

\preprint{LLNL-JRNL-2026180}

\title{
  Stable recurrences of the super-transition array formalism from 
  marked generating functions
}
\author{Daniel \AA{}berg}
\email{aberg2@llnl.gov}
\affiliation{Lawrence Livermore National Laboratory, P.O. Box 808, Livermore, CA 94550, U.S.A.}
\author{Brian G. Wilson}
\affiliation{Lawrence Livermore National Laboratory, P.O. Box 808, Livermore, CA 94550, U.S.A.}
\author{Jean-Christophe Pain}
\affiliation{CEA, DAM, DIF, F-91297 Arpajon, France}
\affiliation{Universit\'e Paris-Saclay, CEA, Laboratoire Mati\`ere en Conditions Extr\^emes,\\
\it 91680 Bruy\`eres-le-Ch\^atel, France}

\begin{abstract} 
The conventional evaluation of super-transition-array (STA) mean energies and variances 
uses partition recurrences with alternating terms that can lose numerical accuracy. 
Here we apply a marked-generating-function approach to derive coupled coefficient 
recurrence relations that avoid these sums and directly accumulate fixed-occupation 
partition functions and arbitrary-order moments of the configuration-average (centroid) 
transition energies. Unlike explicit expansions into products of subshell occupations, 
the resulting recurrences require only one coefficient numerator per moment order. 
The same approach provides recurrences for polynomial occupation-number averages, 
which can be contracted to obtain the contributions of individual configurations 
to the STA variance and the Boltzmann energy correction entering the partition 
function. These contractions avoid the explicit evaluation of individual occupation-product 
averages and reduce the number of auxiliary quantities required. Numerical examples 
demonstrate agreement with direct configuration enumeration, including regimes 
where the alternating Bar--Shalom recurrences lose accuracy. Complementary 
hole-counting recurrences are also derived for all quantities, starting from a full supershell.
\end{abstract}

\maketitle
\section{Introduction}

Super-transition arrays (STAs) were introduced by Bar-Shalom, Oreg, and
Goldstein in 1989~\cite{BarOreGol1989} as a statistical framework for
replacing transition arrays from an impractically large number of
configurations in hot dense plasmas. The original formulation provided
expressions for the first three STA moments: the total strength, mean
transition energy, and variance. Their evaluation relied on recurrence
relations involving alternating sums, both in the construction of the
supershell partition functions and in the subsequent evaluation of the
moments. The cancellation of large positive and negative contributions can lead
to severe numerical instabilities, particularly at low temperatures and
for large supershell occupations. Importantly, accurate partition functions
alone do not eliminate these instabilities, since catastrophic cancellation
can also occur in the expressions for the STA moments themselves.

Later, the same authors extended the approach~\cite{BarOreGol1995} to include
corrections from the unresolved transition arrays (UTAs) associated with
individual configurations. These corrections modify the STA mean energy and
variance through configuration averages of UTA moments derived by Bauche and
coworkers~\cite{BauBauKla1979,BauBauKla1982,BauBauKla1985}.

Oreg, Bar-Shalom, and Klapisch~\cite{OreBarKla1997} subsequently
developed an operator algebra for superconfiguration averages,
showing that polynomials in subshell electron and hole occupations
can be represented by differential operators acting on partition
functions. The resulting averages were expressed in terms of
generalized partition functions with modified degeneracies and
occupations, for which efficient computational recurrences were
also developed.

Wilson and Chen~\cite{WilChe1999} expressed the STA mean and variance
in terms of correlated occupation averages written as generalized 
Fermi factors involving ratios of partition functions, and developed a 
hybrid algorithm for evaluating these ratios. Gilleron and Pain~\cite{GilPai2004} 
introduced a doubly recursive partition-function algorithm over electron 
number and subshell degeneracies that sums only non-negative terms. Together, 
the occupation-number formulation and more stable methods for evaluating 
the required partition functions enabled more robust calculations of STA moments, 
although these could become computationally expensive when many reduced-degeneracy 
partition functions were needed.

Two enhancements were subsequently proposed~\cite{WilGilPai2007}. The first
extended the recurrence relations to count holes rather than electrons when a
supershell is more than half filled. The second reduced the computational cost
of evaluating occupation averages by precomputing and storing selected partition
functions, from which additional reduced-degeneracy partition functions could
be constructed using generating functions. This approach was subsequently clarified
and optimized by Pain, Gilleron, and Wilson~\cite{PaiGilWil2020}, who showed that
the auxiliary coefficients are elementary symmetric polynomials, for which numerous
efficient algorithms are available.

More recently, Pain and Wilson~\cite{PaiWil2024} examined the numerical
difficulties associated with partition functions of reduced degeneracies and
discussed methods for their stable evaluation. They also derived expressions
for STA moments of arbitrary order using multinomial expansions and Stirling
numbers of the second kind, neglecting internal UTA widths in this generalization. 
These expressions avoid alternating reduced-partition sums, but retain explicit 
expansions in products of subshell occupations, which can become combinatorially 
extensive at higher moment orders. Pain and Wilson also noted the desirability 
of extending the stable recursive construction used for partition functions 
to the moments themselves.

In this work, we extend established partition-function coefficient passes to 
closed recurrence systems for the statistical quantities entering STA calculations, 
including centroid-energy moments, internal UTA-width contributions, and 
superconfigurational energy corrections. The approach is based on marked generating 
functions, whose coefficient recurrences allow the required observables to be 
accumulated directly at fixed supershell occupation, without resorting to 
alternating reduced-partition expressions or explicitly evaluating all 
occupation-product averages. We also show how the resulting supershell moments 
can be assembled into arbitrary-order centroid-energy moments of complete STAs.

For the centroid-energy moments, the linear combination of subshell 
occupations defining the transition energy is incorporated directly into 
the generating function. This yields coupled recurrences for moments of 
arbitrary order, requiring only one coefficient numerator per moment order. 
The same construction provides recurrences for general polynomial occupation 
averages. In particular, the electron--hole products entering the internal 
UTA-width contribution and the ordered electron-pair products entering the 
superconfigurational energy correction can be contracted with their respective 
physical coefficients. The resulting recurrences evaluate these contributions 
directly, without propagating all individual occupation-product numerators. 
Finally, we derive complementary hole-counting recurrences for all these 
quantities, providing a complete formulation starting from either an empty 
or a completely filled supershell.

Numerical tests consider a deliberately difficult single supershell and compare
the local partition function, shifted energy, UTA-augmented variance, and
pair-energy contribution with direct enumeration. At one temperature,
selecting the appropriate Bar--Shalom direction gives accurate values across
the occupation range. At a lower temperature, combining the two alternating
directions is insufficient, whereas the electron and hole coefficient passes
agree with the enumerated results.

The paper is organized as follows. Section~\ref{sec:Background} reviews the
STA formalism and notation. Section~\ref{sec:pass_part} develops the
fixed-occupation partition-function pass, and Sec.~\ref{sec:STA_pass}
extends this construction to STA energy moments using a marked generating
function. Section~\ref{sec:occupation_moments} develops the corresponding
recurrences for general occupation polynomials. Sections~\ref{sec:uta_corrections}
and \ref{sec:superconfiguration_energy_correction} derive the contracted
recurrences for the UTA-width and superconfigurational energy corrections,
respectively. Numerical results are presented in Sec.~\ref{sec:results},
followed by discussion and conclusions. The appendices provide the
arbitrary-order assembly of STA moments, the electron--hole formulation,
and the earlier alternating-sign recurrence for the superconfigurational energy correction.

\section{Background and notation}
\label{sec:Background}

\subsection{Configurations, supershells, and fixed-occupancy partitions}

A configuration $ C $ is specified by subshell occupation numbers $ q_s $ with
$ 0 \le q_s \le g_s $, where $ g_s $ is the degeneracy of subshell $ s $. A
supershell $ \sigma $ is a grouping of subshells, and a fixed set of supershell
occupancies $ \{ Q_\sigma \} $ defines a superconfiguration $ \Xi $ through the
constraints
\begin{align}
  \sum_{s\in\sigma} q_s = Q_\sigma
  \qquad
  \text{for each supershell } \sigma .
  \label{eq:supershell-occupation-constraint}
\end{align}
With $ \mathbf{g}_\sigma = \{ g_s \}_{s\in\sigma} $ denoting the degeneracies in
supershell $ \sigma $, following Bar-Shalom et al.~\cite{BarOreGol1989}, the
canonical supershell partition function at fixed occupation is
\begin{align}
  U_{Q_\sigma}( \mathbf{g}_\sigma )
  &=
  \sum_{|\mathbf q_\sigma|=Q_\sigma}
  \prod_{s\in\sigma}
  \binom{g_s}{q_s} X_s^{\,q_s},
  \label{eq:UQ-def}
\end{align}
where
\begin{align}
  X_s = \exp[-\beta(u_s-\mu)].
\end{align}
Here $ u_s $ denotes the one-particle reference energy associated with
subshell $ s $, and denotes a sum over all allowed subshell occupations 
within supershell $\sigma$. Equivalently,
\begin{align}
  U_{Q_\sigma}( \mathbf{g}_\sigma )
  =
  [ z^{Q_\sigma} ]
  \prod_{s\in\sigma}
  ( 1 + X_s z )^{g_s}.
\end{align}
Here $ [ z^Q ] f(z) $ denotes the coefficient of $ z^Q $ in the expansion of
$ f(z) $. For any function $ F(\{ q_s \}) $ of the subshell occupations within
supershell $ \sigma $, we denote its fixed-occupation average by
\begin{align}
  \big\langle F(\{ q_s \}) \big\rangle_\sigma
  =
  \frac{
    \displaystyle
    \sum_{|\mathbf q_\sigma|=Q_\sigma}
    F(\{ q_s \})
    \prod_{s\in\sigma}
    \binom{g_s}{q_s} X_s^{\,q_s}
  }{
    U_{Q_\sigma}( \mathbf{g}_\sigma )
  }.
  \label{eq:canon-avg}
\end{align}

\subsection{Configuration-average transition energy}

For a one-electron transition $ \alpha \to \beta $, let $ C' $ be the
configuration obtained from $ C $ by removing one electron from subshell
$ \alpha $ and placing it in subshell $ \beta $. Following
Refs.~\cite{BarOreGol1989,BarOreGol1995}, the configuration-average transition
energy may be written as
\begin{align}
  E_{\alpha\beta}^{CC'}
  =
  D_0^{\alpha\beta}
  +
  \sum_s q_s^\alpha D_s^{\alpha\beta},
  \label{eq:CfgAvgEnergy}
\end{align}
where $ q_s^\alpha = q_s - \delta_{s\alpha} $. The constant and spectator
coefficients are
\begin{align}
  D_0^{\alpha\beta}
  &= \langle \beta \rangle - \langle \alpha \rangle, \\
  D_s^{\alpha\beta}
  &= \langle s,\beta \rangle - \langle s,\alpha \rangle.
\end{align}
The one-body contribution is
\begin{align}
  \langle s \rangle
  =
  \epsilon_s
  +
  \braketBig{s}{ -\frac{Z}{r} - V(r) }{ s },
  \label{eq:e1}
\end{align}
and the two-body contribution between orbitals $ r $ and $ s $ is
\begin{align}
  \langle r,s \rangle
  &=
  F^0(r,s)
  -
  \left[ \frac{g_r}{g_r - \delta_{r,s}} \right] \nonumber \\
  &\times
  \sum_k
  \threej{ j_s }{ k }{ j_r }{ \tfrac{1}{2} }{ 0 }{ -\tfrac{1}{2} }^{\!2}
  G^k(r,s),
  \label{eq:e2e}
\end{align}
with $ F^k $ and $ G^k $ the Slater integrals.

\subsection{Reduced STA measure and supershell moments}

The raw STA moments associated with the transition $ \alpha \to \beta $ are
dipole-weighted averages of powers of $ E_{\alpha\beta}^{CC'} $ over all
configurations $ C \in \Xi $. The dipole weight contains
$ q_\alpha ( g_\beta - q_\beta ) $. Standard binomial-coefficient identities
absorb this factor into reduced spectator occupations and degeneracies,
\begin{align}
  q_s^\alpha
  &=
  q_s - \delta_{s\alpha},
  \\
  g_s^{\alpha\beta}
  &=
  g_s - \delta_{s\alpha} - \delta_{s\beta},
  \\
  Q_\sigma^\alpha
  &=
  Q_\sigma - \mathbf{1}_{\alpha\in\sigma},
\end{align}
After removing the active $\alpha$ electron and the available $\beta$ state,
the remaining sum is an ordinary fixed-occupation spectator problem with
reduced degeneracies. The contribution of supershell $\sigma$ to the
configuration-average transition energy is
\begin{align}
  E_{\alpha\beta}^\sigma
  =
  \sum_{s\in\sigma} q_s^\alpha D_s^{\alpha\beta},
  \label{eq:transition-energy-supershell-sum}
\end{align}
The standard low-order STA decomposition separates the mean and variance into
supershell sums, with the constant term $D_0^{\alpha\beta}$ in the mean
\cite{BarOreGol1989}:
\begin{align}
  E_{\alpha\beta}
  &=
  D_0^{\alpha\beta}
  +
  \sum_{\sigma\in\Xi}
  \mathcal{E}_{\alpha\beta}^\sigma,
  \label{eq:STA-mean}
  \\
  \Delta E_{\alpha\beta}^2
  &=
  \sum_{\sigma\in\Xi}
  \big( \Delta_{\alpha\beta}^\sigma \big)^2.
  \label{eq:STA-variance}
\end{align}
The corresponding supershell quantities may be written in terms of the
unnormalized raw-moment numerators 
\begin{align}
  A_{m,\sigma}(Q_\sigma^\alpha)
  &=
  \sum_{|\mathbf q_\sigma|=Q_\sigma^\alpha} 
  \big( E_{\alpha\beta}^\sigma \big)^m
  \prod_{s\in\sigma}
  \binom{g_s^{\alpha\beta}}{q_s^\alpha}
  X_s^{\,q_s^\alpha}.
  \label{eq:supershell-average-direct}
\end{align}
In particular,
\begin{align}
  A_{0,\sigma}(Q_\sigma^\alpha)
  =
  U_{Q_\sigma^\alpha}(\mathbf{g}_\sigma^{\alpha\beta}).
\end{align}
Here $A_{0,\sigma}$ is the reduced supershell partition function,
$A_{1,\sigma}/A_{0,\sigma}$ is the contribution to the mean, and
$A_{2,\sigma}/A_{0,\sigma}$ is the second raw moment:
\begin{align}
  \mathcal{E}_{\alpha\beta}^\sigma
  &=
  \left\langle E_{\alpha\beta}^\sigma \right\rangle_\sigma^{\alpha\beta}
  =
  \frac{A_{1,\sigma}(Q_\sigma^\alpha)}
       {A_{0,\sigma}(Q_\sigma^\alpha)},
  \label{eq:deltaE-sigma}
  \\
  \eta_{\alpha\beta}^\sigma
  &=
  \left\langle \left( E_{\alpha\beta}^\sigma \right)^2
  \right\rangle_\sigma^{\alpha\beta}
  =
  \frac{A_{2,\sigma}(Q_\sigma^\alpha)}
       {A_{0,\sigma}(Q_\sigma^\alpha)},
  \\
  \big( \Delta_{\alpha\beta}^\sigma \big)^2
  &=
  \eta_{\alpha\beta}^\sigma
  -
  \big( \mathcal{E}_{\alpha\beta}^\sigma \big)^2.
  \label{eq:deltaV-sigma}
\end{align}
The corresponding supershell variance is
$(\Delta_{\alpha\beta}^\sigma)^2=A_{2,\sigma}/A_{0,\sigma}
-(A_{1,\sigma}/A_{0,\sigma})^2$. For higher-order moments, the full STA raw
moments are sums of products of supershell raw moments; the all-order
construction is given in
Appendix~\ref{app:moment-factorization}.

\subsection{UTA corrections}

The unresolved-transition-array corrections of
Ref.~\cite{BarOreGol1995} modify the STA moments in two distinct ways. First,
the configuration-array centers shift, which preserves the form of the STA mean
formula provided the spectator coefficients are replaced by
\begin{align}
  D_s^{\prime\,\alpha\beta}
  =
  D_s^{\alpha\beta}
  +
  \left(
    \delta_{s\alpha}
    -
    \frac{j_\alpha}{j_\beta}\delta_{s\beta}
  \right)
  \delta E_{\alpha\beta},
  \label{eq:uta-shifted-D}
\end{align}
where $ \delta E_{\alpha\beta} $ is the UTA shift quantity of
Ref.~\cite{BarOreGol1995}. Second, the finite widths of the unresolved
configuration arrays add a positive contribution to the STA variance. In the
reduced active-electron measure, this width term depends on occupation factors
\begin{align}
  q_s^\alpha \left( g_s^{\alpha\beta} - q_s^\alpha \right),
\end{align}
so its evaluation requires reduced fixed-occupation averages of the form
\begin{align}
  \left\langle
    q_s^\alpha \left( g_s^{\alpha\beta} - q_s^\alpha \right)
  \right\rangle_\sigma^{\alpha\beta}.
  \label{eq:uta-width-factor-review}
\end{align}

\subsection{Superconfigurational energy correction}
\label{sec:background_boltzmann}
The supershell partition factors are built from the independent-particle
reference energies $ u_s $ used in $ X_s $. Let
\begin{align}
  E_C^0
  =
  \sum_s q_s u_s
  \label{eq:reference-configuration-energy}
\end{align}
denote the corresponding reference configuration energy. For the
Boltzmann-weighted superconfiguration average, the superconfiguration
contribution to the partition function is approximated by
\begin{align}
  U_\Xi
  \approx
  \left[
    \prod_{\sigma\in\Xi}
    U_{Q_\sigma}(\mathbf{g}_\sigma)
  \right]
  \exp\!\left[
    -\beta \Delta E_\Xi
  \right],
  \label{eq:superconfiguration-partition-correction}
\end{align}
where
\begin{align}
  \Delta E_\Xi
  =
  \left\langle E_C - E_C^0 \right\rangle_{0,\Xi}
  \label{eq:superconfiguration-energy-correction}
\end{align}
is the canonical average over all configurations in $ \Xi $ with weight
$ \prod_r \binom{g_r}{q_r} X_r^{\,q_r} $. Using the configuration-average
energy decomposition of Sec.~\ref{sec:Background}, the correction may be
written as
\begin{align}
  \Delta E_\Xi
  &=
  \sum_{\sigma\in\Xi}
  \sum_{s\in\sigma}
  \langle q_s \rangle_\sigma
  \left( \langle s \rangle - u_s \right)
  \nonumber \\
  &\quad
  +
  \frac{1}{2}
  \sum_{\sigma\ne\tau}
  \sum_{\substack{s\in\sigma \\ t\in\tau}}
  \langle q_s \rangle_\sigma
  \langle q_t \rangle_\tau
  \langle s,t \rangle
  \nonumber \\
  &\quad
  +
  \frac{1}{2}
  \sum_{\sigma\in\Xi}
  \sum_{s,t\in\sigma}
  \left\langle
    q_s \left( q_t - \delta_{st} \right)
  \right\rangle_\sigma
  \langle s,t \rangle.
  \label{eq:thermal-boltzmann-correction-review}
\end{align}
Bar-Shalom's original first-order configuration-energy correction is
statistical~\cite{BarOreGol1989}, whereas the quantity considered here is the
corresponding Boltzmann-weighted superconfigurational energy correction
entering the superconfiguration Boltzmann factor. Bar-Shalom later derived
empty- and full-supershell recurrences for this Boltzmann-weighted first-order
correction in unpublished notes~\cite{BarShalom2003Notes}. The corresponding
recurrences are given in
Appendix~\ref{app:unstable-boltzmann}.

\section{Stable pass for supershell partition functions}
\label{sec:pass_part}

The supershell partition function for $Q_\sigma$ electrons in supershell $\sigma$, 
Eq.~\eqref{eq:UQ-def}, can be obtained as the coefficient of $z^{Q_\sigma}$ from the 
supershell generating function,
\begin{align}
  f_\sigma(z)
  &=
  \prod_{s\in\sigma} \big( 1 + z X_s\big)^{ g_s }, \\
U_{\sigma}(Q_\sigma) &= [ z^{Q_\sigma} ] \, f_\sigma(z).
  \label{eq:genfun}
\end{align}
The coefficient can be built by repeated multiplication by a single factor
$(1 + z X_s)$ starting from 1. If we let
\begin{align}
  f^{(n)}(z) = \sum_Q c_Q^{(n)} z^Q,
  \label{eq:An}
\end{align}
denote the partial generating product after $n$ single-state folds, then folding
one state of subshell $s$ gives
\begin{align}
  f^{(n+1)}(z)
  =
  \sum_Q c_Q^{(n)} z^Q + X_s \sum_Q c_Q^{(n)} z^{Q+1}.
  \label{eq:Anp1}
\end{align}
Equating powers of $z$, we obtain the coefficient relation
\begin{align}  
c_Q^{(n+1)} = c_Q^{(n)} + X_s c_{Q-1}^{(n)},
\end{align} 
which in terms of the partition function, gives the following recurrence relation
\begin{align}
  U^{(n+1)}_\sigma(Q)  &=  U^{(n)}_\sigma(Q) + X_s U^{(n)}_\sigma(Q-1),
  \label{eq:pass-U}
\end{align}
for $ 1 \leq Q \leq Q_\sigma$. So, starting from the empty product, we initialize the 
recurrence by
\begin{align}
  U^{(0)}_\sigma(Q) = \delta_{Q,0},
  \label{eq:init-U}
\end{align}
and apply the update $g_s$ times for each $s\in\sigma$. Because the right-hand side
uses coefficients from the preceding iteration, the update may be performed in
place by traversing $Q$ in descending order. 

We note that Eq.~\eqref{eq:pass-U} is equivalent to the second of the two recurrence 
relations for partition functions derived by Bar-Shalom \emph{et al.} \cite{BarOreGol1989}. 
Their Eq.~(45) reads
$U_Q(\mathbf{g}) = U_Q(\mathbf{g}^\gamma) + X_\gamma U_{Q-1}(\mathbf{g}^\gamma)$,
where $\mathbf{g}^\gamma$ is obtained by reducing the degeneracy of subshell $\gamma$
by one. Although derived there as an identity between partition functions with different
statistical weights, the formulation above makes explicit how this relation can be evaluated 
as a sequence of stable, in-place coefficient updates, starting from the empty product.

The $g_s$ successive single-state updates for a given subshell $s$ can also be combined 
into a single step. If we again assume we have $f^{(n)}(z) = \sum_Q c_Q^{(n)} z^Q$ after 
completely folding in a certain number of subshells, and then multiplying by $(1+zX_s)^{g_s}$
and equating powers of $z$, we get 
\begin{align}
c^{(n+g_s)}_Q &= \sum_{k=0}^{g_s} \binom{g_s}{k} X_s^{k} c_{Q-k}^{(n)}.
\label{eq:coeff1_for_GS}
\end{align}
As a result, we get the following update for the partition function
\begin{align}
U^{\mathrm{new}}_\sigma(Q)
= 
\sum_{k=0}^{\min(g_s,Q)}
\binom{g_s}{k}X_s^k
U^{\mathrm{old}}_\sigma(Q-k),
\label{eq:gp-collapsed-subshell}
\end{align}
which is nothing else than the stable recurrence derived by Gilleron and Pain \cite{GilPai2004}, Eq.~(25). 
Although the subshell-level recurrence processes an entire subshell at once, each 
update requires a convolution over subshell occupations, including binomial coefficients 
and powers of $X_s$. The single-state recurrence instead requires only additions and 
multiplications by $X_s$, permits in-place updates without explicitly evaluating 
binomial coefficients, and provides a convenient foundation for the marked generating 
functions considered below.

\section{Stable passes for STA moments}
\label{sec:STA_pass}
We now extend the generating-function construction of Sec.~\ref{sec:pass_part} to 
obtain stable recurrence relations for the STA energy moments alongside the supershell 
partition function. Specifically, we derive recurrences for the unnormalized raw-moment 
numerators $A_{m,\sigma}(Q_\sigma^\alpha)$ introduced in 
Eq.~\eqref{eq:supershell-average-direct}. 
These determine the supershell contributions to the moments of the configuration-average 
transition energies. The same approach will subsequently be extended to include UTA shifts 
and widths in Sec.~\ref{sec:uta_corrections}.

For simplicity, we suppress the explicit transition and supershell labels throughout 
this section and write $q_s=q_s^\alpha$, $Q=Q_\sigma^\alpha$, $g_s=g_s^{\alpha\beta}$, 
and $D_s=D_s^{\alpha\beta}$. Thus, $q_s$ and $g_s$ denote the reduced occupations and 
degeneracies.

The supershell contribution to the configuration-average transition energy is, 
from Eq.~\eqref{eq:transition-energy-supershell-sum},
\begin{align}
  E(\mathbf q) = \sum_{s\in\sigma} q_s D_s.
\end{align}
To generate moments of this quantity, we introduce an auxiliary variable $t$ into 
the partition-function generating product. In Sec.~\ref{sec:pass_part}, each 
occupied spin-orbital contributes a factor $zX_s$ when folded in. Multiplying 
this factor by $e^{tD_s}$ assigns an additional weight $e^{tq_sD_s}$ to a subshell 
containing $q_s$ electrons, and therefore a weight $e^{tE(\mathbf q)}$ to the entire 
configuration. The corresponding {\em marked} generating function is then 
\begin{align}
  \mathcal{F}_\sigma(z,t)
  =
  \prod_{s\in\sigma}
  \left(1 + z X_s e^{tD_s}\right)^{g_s}.
  \label{eq:marked-supershell-generating-function}
\end{align}

Extracting the coefficient of $z^Q$ selects the fixed 
supershell occupation, while differentiating with respect 
to $t$ and evaluating at $t=0$ 
generates the powers of the supershell energy contribution. 
More precisely, we have
\begin{align}
  \mathcal{F}_\sigma(z,t)
  &=
  \sum_{\{q_\sigma\}}
  \prod_{s\in\sigma}
  \binom{g_s}{q_s}\left(z X_s e^{t D_s}\right)^{q_s} \\
  &= \sum_{\{q_\sigma\}} z^{\sum_r q_r}e^{t\sum_r q_r D_r}
  \prod_{s\in\sigma}
  \binom{g_s}{q_s} X_s^{q_s},
\end{align} 
so extracting coefficient for $z^Q$ gives 
\begin{align}
  [z^Q]\mathcal{F}_\sigma(z,t)
  =
  \sum_{|\mathbf q|=Q}
  e^{tE_\sigma(\mathbf q)}
  \prod_{s\in\sigma}
  \binom{g_s}{q_s}X_s^{q_s}.
  \label{eq:local-mgf-coefficient}
\end{align}
This coefficient is the unnormalized fixed-occupation moment-generating
function, and its derivatives define the unnormalized moment numerators,
\begin{align}
  A_m(Q)
  =
  \left.
  \frac{\partial^m}{\partial t^m}
  [z^Q]\mathcal{F}_\sigma(z,t)
  \right|_{t=0},
  \label{eq:Am-from-mgf}
\end{align}
with $A_0(Q)=U(Q)$. 

To derive recurrence relations for the moments, 
let $f^{(n)}(z,t)$ be the partial product after $n$ one-electron states have 
been included and define
\begin{align}
  f^{(n)}(z,t)
  &=
  \sum_Q F_Q^{(n)}(t)z^Q,
\end{align}
with 
\begin{align}
  A_m^{(n)}(Q)
  &=
  \left.
  \frac{\partial^m F_Q^{(n)}(t)}{\partial t^m}
  \right|_{t=0}.
\end{align}
Including one further available state of subshell $s$ gives
\begin{align}
  f^{(n+1)}(z,t)
  &=
  \left(1+zX_s e^{tD_s}\right)f^{(n)}(z,t),
  \label{eq:moment_one_update}
\end{align}
and equating powers of $z$ leads to 
\begin{align}
  F_Q^{(n+1)}(t)
  &=
  F_Q^{(n)}(t)+X_s e^{tD_s}F_{Q-1}^{(n)}(t).
  \label{eq:marked-single-fold}
\end{align}
Differentiating the coefficient recurrence in
Eq.~\eqref{eq:marked-single-fold} $m$ times with respect to $t$ and 
applying the product rule yields
\begin{align}
  \frac{\partial^m}{\partial t^m}
  \left[e^{tD_s}F_{Q-1}^{(n)}(t)\right]
  =
  \sum_{j=0}^{m}\binom{m}{j}D_s^{m-j}
  \frac{d^jF_{Q-1}^{(n)}(t)}{dt^j}.
\end{align}
Evaluating at $t=0$ then gives the final recurrence relation for 
the STA raw moment numerators
\begin{align}
  A_m^{(n+1)}(Q)
  &=
  A_m^{(n)}(Q) \nonumber \\
  & \quad + X_s\sum_{j=0}^{m}\binom{m}{j}D_s^{m-j}A_j^{(n)}(Q-1).
  \label{eq:all-order-moment-pass}
\end{align}
The recurrence is closed at each moment order since the update of $A_m$ 
depends only on $A_j$ with $j\leq m$. As a result, raw moments of any order 
can be obtained by propagating a finite set of numerators.
For $m=0,1,2$, we obtain
\begin{widetext}
\begin{align}
  A_0^{(n+1)}(Q)
  &=A_0^{(n)}(Q)+X_sA_0^{(n)}(Q-1),
  \label{eq:a0-pass}
  \\
  A_1^{(n+1)}(Q)
  &=A_1^{(n)}(Q)+X_s\Big[A_1^{(n)}(Q-1)+D_sA_0^{(n)}(Q-1)\Big],
  \label{eq:a1-pass}
  \\
  A_2^{(n+1)}(Q)
  &=A_2^{(n)}(Q)
  +X_s\Big[A_2^{(n)}(Q-1)+2D_sA_1^{(n)}(Q-1)
  +D_s^2A_0^{(n)}(Q-1)\Big].
  \label{eq:a2-pass}
\end{align}
\end{widetext}
Since $A_0 = U$ the same pass yields the partition function and the first two raw-moment numerators.
Before any one-electron states have been included, the generating 
product is $f^{(0)}(z,t)=1$, and hence the recurrences are initialized by
\begin{align}
A_m^{(0)}(Q) = \delta_{Q0} \delta_{m0},
\end{align}
for non-negative $m$.

After all spin-orbitals have been incorporated, the numerators yield the supershell 
energy moments. Restoring the transition and supershell labels, the mean and variance are
\begin{align}
  \mathcal{E}_{\alpha\beta}^\sigma
  &=
  \frac{A_{1,\sigma}(Q_\sigma^\alpha)}
       {A_{0,\sigma}(Q_\sigma^\alpha)},\\
  \big(\Delta_{\alpha\beta}^\sigma\big)^2
  &=
  \eta_{\alpha\beta}^\sigma
  -\big(\mathcal{E}_{\alpha\beta}^\sigma\big)^2,
  \label{eq:local-moments-from-A}
\end{align}
where the second raw moment is
\begin{align}
  \eta_{\alpha\beta}^\sigma
  &=
  \frac{A_{2,\sigma}(Q_\sigma^\alpha)}
       {A_{0,\sigma}(Q_\sigma^\alpha)}.
\end{align}
 Higher raw moments follow directly from Eq.~\eqref{eq:all-order-moment-pass}, while 
the construction of arbitrary-order central moments is detailed in 
Appendix~\ref{app:moment-factorization}.

The recurrences above construct the generating product by successively 
including single-electron states, starting from the empty product. An equivalent 
hole formulation, in which the fully occupied supershell serves as the reference 
and holes are introduced successively, is derived in Appendix~\ref{app:electron-hole-passes}.

Alternatively, an entire subshell can be incorporated at once by multiplying 
the generating function by $(1+zX_s e^{tD_s})^{g_s}$, giving
\begin{align}
F_Q^{(n+g_s)}(t)
&=
\sum_{k=0}^{\min(g_s,Q)}
\binom{g_s}{k}X_s^k e^{tkD_s}
F_{Q-k}^{(n)}(t).
\end{align}
Differentiation with respect to $t$ yields coupled moment recurrences 
generalizing the Gilleron--Pain partition-function update, 
Eq.~\eqref{eq:gp-collapsed-subshell}. We retain the single-spin-orbital 
formulation because it also provides a convenient starting point for 
the occupation-moment recurrences developed in the following section.
\section{Subshell occupation moments}
\label{sec:occupation_moments}

We now extend the marked generating-function formalism of the preceding
section to obtain recurrence relations for moments and polynomial functions
of subshell occupations. Such quantities enter the UTA widths and
superconfiguration energy corrections, while average subshell occupations
are also needed in self-consistent-field calculations of
superconfiguration-dependent wave functions, energies, and cross sections.

We consider a supershell $\sigma$ containing $Q$ electrons, with subshell
occupations $q_s$ and degeneracies $g_s$, where $0\leq q_s\leq g_s$ and
$\sum_{s\in\sigma}q_s=Q$. Unlike the preceding section, $q_s$ and $g_s$
may represent either reduced or unreduced occupations and degeneracies.

For a general polynomial $f(\mathbf q)$ in the subshell occupations,
the supershell average is
\begin{align}
  \langle f\rangle_Q &= \frac{N_f(Q)}{U(Q)},
\end{align}
where the unnormalized numerator is
\begin{align}
  N_f(Q)
  &=
  \sum_{|\mathbf q|=Q}
  f(\mathbf q)w(\mathbf q),
  \label{eq:occ-poly-target}
\end{align}
with the statistical weights
\begin{align}
  w(\mathbf q)
  &=
  \prod_{s\in\sigma}
  \binom{g_s}{q_s}X_s^{q_s}.
\end{align}

Their constrained sum gives the fixed-occupation supershell
partition function,
\begin{align}
  U(Q)
  &=
  \sum_{|\mathbf q|=Q}
  w(\mathbf q).
  \label{eq:occ-moment-weight}
\end{align}

For example, $f=q_s$ gives the average occupation of subshell $s$,
$f=q_s^2$ gives its second raw moment, and $f=q_sq_r$ gives a mixed
occupation moment for $r\ne s$. More generally, $f(\mathbf q)$ may
be any polynomial in the subshell occupations. We now show how the
corresponding unnormalized numerators can be obtained using recurrence
relations.

\subsection{Generating occupation averages}

To generate occupation moments, we introduce the multivariate marked
generating function
\begin{align}
 \mathcal G_\sigma(z,\mathbf t)
 &=\prod_{s\in\sigma}(1+zX_se^{t_s})^{g_s}.
 \label{eq:occ-poly-generator}
\end{align}
Each subshell $s$ is assigned an independent auxiliary variable $t_s$,
which marks its occupation number. Expanding the product and extracting
the coefficient of $z^Q$ gives
\begin{align}
 [z^Q]\mathcal G_\sigma(z,\mathbf t)
 &=\sum_{|\mathbf q|=Q}w(\mathbf q)e^{\sum_s t_sq_s}.
 \label{eq:occ-poly-expansion}
\end{align}
Differentiation with respect to $t_s$ generates powers of the
corresponding occupation number $q_s$:
\begin{align}
 \frac{\partial^k}{\partial t_s^k}
 e^{\sum_r t_rq_r}
 &=q_s^k e^{\sum_r t_rq_r}.
 \label{eq:occ-poly-derivative-examples}
\end{align}
Setting all auxiliary variables to zero then gives the desired
weighted sum of $q_s^k$. Similarly, differentiation with respect
to two distinct markings gives
\begin{align}
 \left.
 \frac{\partial^2}{\partial t_s\partial t_r}
 e^{\sum_u t_uq_u}
 \right|_{\mathbf t=\mathbf0}
 &=q_sq_r,\qquad r\ne s.
 \label{eq:occ-poly-mixed-derivative}
\end{align}

These examples extend to any polynomial $f(\mathbf q)$. Replacing each occupation $q_s$ by the corresponding derivative $\partial/\partial t_s$ defines the differential operator $f(\boldsymbol{\partial}_t)$, where $\boldsymbol{\partial}_t=(\partial/\partial t_s)_{s\in\sigma}$. The unnormalized numerator of Eq.~\eqref{eq:occ-poly-target} is then
\begin{align}
 N_f(Q)
 &=
 \left.
 [z^Q]f(\boldsymbol{\partial}_t)
 \mathcal G_\sigma(z,\mathbf t)
 \right|_{\mathbf t=\mathbf0}.
 \label{eq:occ-poly-operator-numerator}
\end{align}
For example, the occupation polynomial $f=q_s(g_s-q_s)$ corresponds to the operator
$f(\boldsymbol{\partial}_t)=g_s\partial_{t_s}-\partial_{t_s}^2$.

This differential-operator representation is closely related to the
partition-function algebra of Oreg, Bar-Shalom, and
Klapisch~\cite{OreBarKla1997}, who introduced occupation operators
$\hat q_s=X_s\partial/\partial X_s$ and
$\hat h_s=g_s-\hat q_s$ to express polynomial occupation averages
in terms of generalized partition functions with modified degeneracies
and occupations.

Differentiation with respect to the markings $t_s$ is mathematically
equivalent to the action of $\hat q_s$ on the unmarked generating
function. The same occupation recurrences could therefore be obtained
by applying these operators directly to the unmarked coefficient
recurrence. The markings, however, provide a common moment-generating
representation for the centroid-energy and occupation moments, with
the observables encoded directly in the generating product.

In the following subsection, we apply the marking derivatives to
the coefficient recurrence for the partial generating product to
obtain closed recurrences for the required occupation numerators,
without constructing the corresponding reduced-degeneracy
partition functions.

\subsection{Recurrences for occupation averages}

The numerators can be calculated by the same coefficient extraction method 
outlined in Sec.~\ref{sec:STA_pass}. Each spin-orbital in subshell $a$ contributes
a factor $1+zX_ae^{t_a}$ to the generating function; the complete subshell
contains $g_a$ such factors. After $n$ spin-orbitals have been included, 
we write the partial generating product as
\begin{align}
 g^{(n)}(z,\mathbf t)
   =\sum_Q C_Q^{(n)}(\mathbf t)z^Q.
 \label{eq:occ-poly-partial-expansion}
\end{align}
The coefficients $C_Q^{(n)}(\mathbf t)$ retain their dependence on the 
auxiliary variables, and their derivatives define the intermediate occupation numerators
\begin{align}
 N_f^{(n)}(Q)
   =\left.f(\boldsymbol\partial_t)C_Q^{(n)}(\mathbf t)
      \right|_{\mathbf t=\mathbf0},
\label{eq:occ-poly-partial-numerators}
\end{align} 
with
\begin{align}
 U^{(n)}(Q)=C_Q^{(n)}(\mathbf0).
 \label{eq:UfromC}
\end{align}
Multiplication of $g^{(n)}(z,\mathbf t)$ by the factor $(1+zX_a e^{t_a})$ for the next 
spin-orbital in subshell $a$, and matching coefficients of $z^Q$, now gives
\begin{align}
 C_Q^{(n+1)}(\mathbf t)
  =C_Q^{(n)}(\mathbf t)+X_ae^{t_a}C_{Q-1}^{(n)}(\mathbf t).
 \label{eq:occ-poly-coefficient-recurrence}
\end{align}
To obtain a recurrence for the numerator associated with \(f(\mathbf q)\), we 
apply \(f(\boldsymbol{\partial}_t)\) to both sides of this equation and evaluate 
at \(\mathbf t=\mathbf0\).

Consider for example the $k$-th occupation-number moment of subshell $r$, for which
$f(\mathbf q)=q_r^k$. We denote the occupation numerator
by $O_{k,r}=N_{q_r^k}$. At stage $n$ we have 
\begin{align}
 O_{k,r}^{(n)}(Q)
  &=\left.\frac{\partial^k}{\partial t_r^k} C_Q^{(n)}(\mathbf t)\right|_{\mathbf t=\mathbf0},
 \label{eq:occ-poly-partial-occupation-numerators}
\end{align}
which leads to the following relations
\begin{align}
 \frac{\partial^k}{\partial t_r^k}C_Q^{(n+1)}  &= \frac{\partial^k}{\partial t_r^k}C_Q^{(n)}  +
 X_a e^{t_a} \bigg[ \frac{\partial^k}{\partial t_r^k}C_{Q-1}^{(n)}
   \nonumber \\
  &\quad +  \delta_{a,r} \sum_{p=0}^{k-1} \binom{k}{p} \frac{\partial^p}{\partial t_r^p}C_{Q-1}^{(n)}
  \bigg].
\end{align}

Setting $\mathbf{t}=0$ and using Eqs.~(\ref{eq:UfromC}) and 
(\ref{eq:occ-poly-partial-occupation-numerators}) gives the simple recurrence relations 
\begin{align}
O_{k,r}^{(n+1)}(Q)
={}& O_{k,r}^{(n)}(Q)
   + X_a O_{k,r}^{(n)}(Q-1) \nonumber\\
 &+ X_a\delta_{a,r}
 \bigg[ U^{(n)}(Q-1) \nonumber\\
 &\qquad
 + \sum_{p=1}^{k-1}\binom{k}{p}
   O_{p,r}^{(n)}(Q-1) \bigg].
\label{eq:occupation-pass}
\end{align}

The recurrence for $O_{k,r}^{(n+1)}(Q)$ depends only on the partition 
function and occupation numerators of orders $p\leq k$. As a result, 
occupation moments through any prescribed order can be propagated together. 
Starting from the empty generating product $g^{(0)}(z,\mathbf t)=1$, 
the initial conditions are
\begin{align}
 U^{(0)}(Q)&=\delta_{Q0},\\
 O_{k,r}^{(0)}(Q)&=0,\qquad k\geq1.
 \label{eq:occupation-pass-initial}
\end{align}

The same procedure applies to any polynomial of the subshell occupations.
Its differential operator is obtained by replacing each occupation by the
derivative with respect to the corresponding auxiliary variable. Applying
this operator to the coefficient recurrence,
Eq.~\eqref{eq:occ-poly-coefficient-recurrence}, gives
\begin{align}
 f(\boldsymbol\partial_t)C_Q^{(n+1)}
 =f(\boldsymbol\partial_t)C_Q^{(n)}
  +X_af(\boldsymbol\partial_t)\left(e^{t_a}C_{Q-1}^{(n)}\right).
 \label{eq:occ-poly-applied-operator}
\end{align}
Expanding the derivatives with the product rule and setting the auxiliary
variables to zero identifies both the desired numerator and any additional
numerators required for its update. The same procedure can be applied to 
any additional numerators, yielding a closed set of recurrence relations 
involving derivatives of order no higher than that of the original polynomial.

We next apply this construction to two occupation polynomials that enter the 
UTA widths in Sec.~\ref{sec:uta_corrections} and the superconfiguration 
Boltzmann correction in Sec.~\ref{sec:superconfiguration_energy_correction}, respectively.

\subsection{The electron-hole product}
To evaluate the UTA width contribution using the same formalism, we now focus 
on the supershell expectation value of $q_s(g_s - q_s)$, from
which we can form the operator
\begin{align}
    \mathcal{W}_s = \frac{\partial}{\partial t_s}\left( g_s  - \frac{\partial}{\partial t_s}\right).
\end{align} 
We then identify the numerator of the corresponding expectation value at stage $n$ by 
\begin{align}
 W_{s}^{(n)}(Q)
  &=\left.\mathcal{W}_s C_Q^{(n)}(\mathbf t)\right|_{\mathbf t=\mathbf0},
\end{align}
and apply this operator to the coefficient expression to get
\begin{widetext}
\begin{align}
\mathcal{W}_{s}  C_Q^{(n+1)} &= \mathcal{W}_{s}  
\left[
 C_Q^{(n)} + X_a e^{t_a}  C_{Q-1}^{(n)} 
\right] \nonumber \\
&= \mathcal{W}_{s}  C_Q^{(n)} + X_a e^{t_a} 
\biggl[ 
\delta_{a,s}  \biggl( (g_s-1)  C_{Q-1}^{(n)} -  
   2\frac{\partial C^{(n)}_{Q-1}}{\partial t_s}
  \biggr) + \mathcal{W}_{s}  C_{Q-1}^{(n)} 
\biggr].
\end{align}
The recurrence relation for $W_s$ is therefore 
\begin{align}
W_s^{(n+1)}(Q) &=  W_s^{(n)}(Q) + X_a 
\biggl[ \delta_{a,s}  \bigl( (g_s-1)  U^{(n)}(Q-1) -  
   2 O_{1,s} ^{(n)}(Q-1)
  \bigr) + W_s^{(n)}(Q-1)  
\biggr]
\label{eq:w-recurrence}
\end{align}
\end{widetext}
The initial values are $W_s^{(0)}(Q)=0$, since the empty
generating product is independent of the marking variables.
The recurrence therefore closes on $W_s$, $O_{1,s}$,
and $U$, without requiring higher occupation moments.

\subsection{Number of ordered electron pairs}

We next consider the expectation value of
$q_r(q_s-\delta_{rs})$, which counts ordered pairs of distinct
electrons in subshells $r$ and $s$. The corresponding differential
operator is
\begin{align}
  \mathcal M_{rs}
  &=
  \frac{\partial}{\partial t_r}
  \left(
    \frac{\partial}{\partial t_s}-\delta_{rs}
  \right).
\end{align}
The associated numerator at stage $n$ is
\begin{align}
  M_{rs}^{(n)}(Q)
  &=
  \left.
  \mathcal M_{rs} C_Q^{(n)}(\mathbf t)
  \right|_{\mathbf t=\mathbf0}.
\end{align}

Applying this operator to the coefficient recurrence,
Eq.~\eqref{eq:occ-poly-coefficient-recurrence}, gives
\begin{widetext}
\begin{align}
\mathcal M_{rs}C_Q^{(n+1)}
&=
\mathcal M_{rs}
\left[
C_Q^{(n)}+X_a e^{t_a}C_{Q-1}^{(n)}
\right]
\nonumber\\
&=
\mathcal M_{rs}C_Q^{(n)}
+X_a e^{t_a}
\left[
\mathcal M_{rs}C_{Q-1}^{(n)}
+\delta_{as}\frac{\partial C_{Q-1}^{(n)}}{\partial t_r}
+\delta_{ar}\frac{\partial C_{Q-1}^{(n)}}{\partial t_s}
\right].
\end{align}
\end{widetext}
Evaluating at $\mathbf t=\mathbf0$ yields
\begin{align}
M_{rs}^{(n+1)}(Q)
={}&M_{rs}^{(n)}(Q)
+X_a M_{rs}^{(n)}(Q-1)
\nonumber\\
&+X_a\bigl[
\delta_{as}O_{1,r}^{(n)}(Q-1)
\nonumber\\
&\qquad
+\delta_{ar}O_{1,s}^{(n)}(Q-1)
\bigr].
\label{eq:ordered-pair-pass}
\end{align}
The initial values are $M_{rs}^{(0)}(Q)=0$, since the
empty generating product is independent of the marking
variables. The recurrence closes on $M_{rs}$ and the
first occupation numerators $O_{1,r}$ and $O_{1,s}$.
\section{UTA corrections to the STA moments}
\label{sec:uta_corrections}

The STA moments derived in Sec.~\ref{sec:STA_pass} assign a single
transition energy to each configuration. In reality, as each configuration
contains an internal term structure, all transitions between individual 
levels of a pair of configurations can be characterized by the UTA mean transition
energy and variance. 
The UTA mean may differ from the
difference of the configuration-average energies, while its internal
variance accounts for the spread of transition energies within the
array.

As shown by Bar-Shalom, Oreg, and Goldstein
\cite{BarOreGol1995}, these effects modify the STA moments through
a shift of the UTA mean and an additional contribution to the STA
variance. For a fixed transition $\alpha\to\beta$, let
$\overline E_C$ and $v_C$ denote the mean transition energy and
internal variance of the UTA originating from configuration $C$.
The STA mean and variance are then
\begin{align}
 E_{\mathrm{STA}}
 &=\langle\overline E_C\rangle,\\
 V_{\mathrm{STA}}
 &=\langle\overline E_C^{\,2}\rangle
   -\langle\overline E_C\rangle^2
   +\langle v_C\rangle.
 \label{eq:uta-variance-decomposition}
\end{align}
Here, the averages are taken over configurations using the reduced
transition weights introduced earlier.

\subsection{UTA shifts and internal widths}

The UTA shift preserves the linear dependence of the mean transition
energy on the reduced occupations. Equations (A11)--(A12) of
Ref.~\cite{BarOreGol1995} give
\begin{align}
 \overline E_C
 &=D_0^{\alpha\beta}
   +\sum_s q_s^\alpha D_s^{\prime\,\alpha\beta},\\
 D_s^{\prime\,\alpha\beta}
 &=D_s^{\alpha\beta}
   +\left(\delta_{s\alpha}
   -\frac{j_\alpha}{j_\beta}\delta_{s\beta}\right)
   \delta E_{\alpha\beta}.
 \label{eq:uta-shifted-coefficients}
\end{align}
Here $\delta E_{\alpha\beta}$ is independent of the configuration
occupations. The energy-moment recurrences of
Sec.~\ref{sec:STA_pass} therefore apply directly with the shifted
coefficients $D_s^{\prime\,\alpha\beta}$.
The statistical weights, partition function, and
$D_0^{\alpha\beta}$ remain unchanged.

The internal UTA variance is given by Eq.~(A13) of
Ref.~\cite{BarOreGol1995} as
\begin{align}
 v_C
 &=\sum_s(\Delta_s^{\alpha\beta})^2
   (q_s-\delta_{s\alpha})
   (g_s-q_s-\delta_{s\beta}).
 \label{eq:uta-configuration-width-original}
\end{align}
Introducing the reduced occupations and degeneracies,
\begin{align}
 q_s^\alpha&=q_s-\delta_{s\alpha},\\
 g_s^{\alpha\beta}
 &=g_s-\delta_{s\alpha}-\delta_{s\beta},
\end{align}
we obtain
\begin{align}
 v_C
 &=\sum_s(\Delta_s^{\alpha\beta})^2
   q_s^\alpha
   (g_s^{\alpha\beta}-q_s^\alpha).
 \label{eq:uta-configuration-width-reduced}
\end{align}

For a single supershell, we again suppress the explicit transition
and supershell labels and write
\begin{align}
 q_s&\equiv q_s^\alpha,\quad
 g_s\equiv g_s^{\alpha\beta},\quad
 Q\equiv Q_\sigma^\alpha,\nonumber\\
 \Delta_s^2&\equiv(\Delta_s^{\alpha\beta})^2.
\end{align}
The electron-hole numerator derived in
Sec.~\ref{sec:occupation_moments} is
\begin{align}
 W_s(Q)
 &=\sum_{\sum_r q_r=Q}
   q_s(g_s-q_s)w(\mathbf q).
 \label{eq:uta-width-numerator-target}
\end{align}
Consequently, the contribution of supershell $\sigma$ to the
average internal UTA variance is
\begin{align}
 \langle v_\sigma\rangle_Q
 &=\frac{1}{U(Q)}
   \sum_{s\in\sigma}\Delta_s^2 W_s(Q).
 \label{eq:uta-supershell-width-average}
\end{align}

\subsection{Contraction of the UTA width}
\label{sec:uta-contraction}
In evaluating Eq.~(\ref{eq:uta-supershell-width-average}), it turns out it is 
not necessary to evaluate $W_s(Q)$ individually for each $s$ in $\sigma$. 
Rather, one can directly obtain a recurrence relation for the numerator
\begin{align}
L(Q) = \sum_s  \Delta_s^2 W_s(Q)
\end{align}
by writing
\begin{align*}
L^{(n+1)}(Q) &= \sum_s \Delta_s^2 W^{(n+1)}_s(Q),
\end{align*}
inserting the recurrence relation of Eq.~(\ref{eq:w-recurrence}), and simplifying
the summation using the $\delta$-functions to give
\begin{widetext}
\begin{align}
L^{(n+1)}(Q) &= L^{(n)}(Q) + 
 X_a 
\biggl[  \Delta_a^2 \bigl( (g_a-1)  U^{(n)}(Q-1) -  
   2 O_{1,a} ^{(n)}(Q-1)
  \bigr) + L^{(n)}(Q-1) 
\biggr].
\end{align}
\end{widetext}
By defining 
\begin{align}
      V_a^{(n)}(Q) = (g_a-1)  U^{(n)}(Q) - 2 O_{1,a}^{(n)}(Q),
\end{align}
the expression for $L$ simplifies to 
\begin{align}
L^{(n+1)}&(Q) = L^{(n)}(Q) \nonumber \\
& + 
 X_a 
\left[ \Delta_a^2 V^{(n)}_a(Q-1) + L^{(n)}(Q-1) 
\right],
\label{eq:rec_L} 
\end{align}
with $L^{(0)}(Q) = 0$. We can easily derive a recurrence relation 
for $V$ by writing 
\begin{align}
V_t^{(n+1)}(Q) &= (g_t-1)  U^{(n+1)}(Q) - 2 O_{1,t}^{(n+1)}(Q),
\end{align}
and make use of Eq.~(\ref{eq:occupation-pass}) to give
\begin{align}
V_t&^{(n+1)}(Q) 
= V_t^{(n)}(Q) \nonumber \\
&+ X_a  \left[V_t^{(n)}(Q-1) -2 \delta_{a,t}
 U^{(n)}(Q-1) \right],
\end{align}
with $V_t^{(0)}(Q)=(g_t-1)\delta_{Q0}$ for all subshells $t\in\sigma$. 

At first sight, the recurrence for $V_t$ appears to require a separate 
numerator for every subshell. However, by folding all spin-orbitals of 
each subshell consecutively, only the numerator associated with the current 
subshell is needed. Suppose that all preceding subshells have been 
incorporated and we are about to begin subshell $b$. Since none of its 
spin-orbitals have yet been included, $O_{1,b}(Q)=0$, and hence
\begin{align}
V_b(Q)=(g_b-1)U(Q).
\end{align}
During the subsequent $g_b$ folds, the recurrence simplifies to
\begin{align}
V_b^{(n+1)}&(Q)
=V_b^{(n)}(Q)\nonumber\\
&+X_b\left[
V_b^{(n)}(Q-1)-2U^{(n)}(Q-1)
\right].
\end{align}
Once all spin-orbitals of subshell $b$ have been incorporated, its 
auxiliary numerator $V_b$ is no longer needed. The calculation therefore 
requires only the partition-function numerator $U(Q)$, the contracted 
width numerator $L(Q)$, and one temporary auxiliary numerator $V_b(Q)$, 
which is reinitialized at the beginning of each subshell.

After all spin-orbitals have been incorporated, the average internal UTA variance 
in Eq.~(\ref{eq:uta-supershell-width-average}) is obtained directly as 
$\langle v_\sigma\rangle_Q=L(Q)/U(Q)$.
\section{Evaluation of the superconfigurational energy correction}
\label{sec:superconfiguration_energy_correction}

We now turn to the evaluation of the Boltzmann-weighted
superconfigurational energy correction introduced in
Sec.~\ref{sec:background_boltzmann}. As shown in
Eq.~\eqref{eq:thermal-boltzmann-correction-review},
the one-body contribution requires only the mean subshell
occupations, which can be obtained using the recurrences
of Sec.~\ref{sec:occupation_moments}. The same is true for
interactions between electrons in different supershells,
whose occupation averages factorize under the reference
statistical measure.

The remaining contribution involves pairs of electrons
within the same supershell. Because the total supershell
occupation is fixed, the individual subshell occupations
are correlated, and their pair averages must be evaluated
together. We should also note that all occupations and 
degeneracies appearing in this section
are unreduced.

\subsection{Pair-energy numerator}

For one supershell $\sigma$, the contribution from the last term of
Eq.~\eqref{eq:thermal-boltzmann-correction-review} is
\begin{align}
 P_\sigma=\frac12\sum_{s,u\in\sigma}\langle s,u\rangle
       \left\langle q_s(q_u-\delta_{su})\right\rangle_{Q_\sigma}.
 \label{eq:sc-energy-pair-average}
\end{align}
The factor $q_s(q_u-\delta_{su})$ counts ordered pairs of distinct
electrons, while the factor $1/2$ removes the double counting
in the sum.

In the notation of Sec.~\ref{sec:occupation_moments}, we define
the corresponding occupation polynomial
\begin{align}
 p(\mathbf q)=\frac12\sum_{r,s\in\sigma}\langle r,s\rangle
                     q_r(q_s-\delta_{rs}),
\end{align}
and its unnormalized numerator
\begin{align}
 K_\sigma(Q)=N_p(Q)
  =\sum_{\sum_s q_s=Q}w(\mathbf q)p(\mathbf q).
 \label{eq:sc-energy-pair-numerator}
\end{align}
From Eq.~(\ref{eq:ordered-pair-pass}), the required pair-energy average is then
\begin{align}
 P_\sigma &=\frac{K_\sigma(Q_\sigma)}{U_\sigma(Q_\sigma)} 
 \label{eq:sc-energy-pair-normalization}
\end{align}

\subsection{Contraction of the pair energy}
\label{sec:boltzmann-contraction}
As in Sec.~\ref{sec:uta-contraction}, we can avoid evaluating the individual pair-occupation numerators. 
For a supershell containing $n_\sigma$ subshells, this eliminates the need to 
propagate $n_\sigma(n_\sigma+1)/2$ independent numerators $M_{rs}$ in 
\begin{align}
K_\sigma(Q_\sigma) &= \frac{1}{2} \sum_{r,s\in\sigma}\langle r,s\rangle M_{rs}(Q_\sigma).
\end{align}
By temporarily dropping the subscript $\sigma$, using Eq.~(\ref{eq:ordered-pair-pass}), 
and the fact that $\langle r,s\rangle = \langle s,r\rangle$, 
we can derive a recurrence relation directly for $K$ by
\begin{align}
K^{(n+1)}(Q) &= \frac12 \sum_{r,s}\langle r,s\rangle M^{(n+1)}_{rs}(Q) \nonumber \\
&= K^{(n)}(Q) + X_a \biggl[ K^{(n)}(Q-1) \nonumber \\
  &\quad +  \sum_{r} \langle a,r\rangle O_{1,r}^{(n)}(Q-1)
 \biggr].
\end{align}
This suggests defining the auxiliary numerator $J_a$ by 
\begin{align}
J_a^{(n)}(Q) &=  \sum_{u} \langle a,u\rangle O_{1,u}^{(n)}(Q),
\end{align}
such that we can write 
\begin{align}
K^{(n+1)}&(Q) = K^{(n)}(Q)  \nonumber \\
  &+ X_a \left( K^{(n)}(Q-1) + J_a^{(n)}(Q-1)
 \right),
\end{align}
with $K^{(0)}(Q)=0$. The recurrence relation for the auxiliary
quantities $J_a$ is easily derived from Eq.~(\ref{eq:occupation-pass}) to be 
\begin{align}
J_b^{(n+1)}(Q) &=  \sum_{u} \langle b,u \rangle O_{1,u}^{(n+1)}(Q) \nonumber \\
  &= J_b^{(n)}(Q)  + X_a \biggl[ J_b^{(n)}(Q-1) \nonumber \\
  &\quad  + \langle a,b \rangle U^{(n)}(Q-1) \biggr].
\end{align} 
Since $O_{1,r}^{(0)}(Q)=0$, the initial values are
$J_b^{(0)}(Q)=0$ for all $b$. Unlike the temporary auxiliary
numerator $V_b(Q)$ of Sec.~\ref{sec:uta-contraction}, the
inhomogeneous term in the recurrence for $J_b$ does not vanish
when $a\ne b$. Consequently, each $J_b$ accumulates contributions
from previously incorporated subshells, and all $n_\sigma$
auxiliary numerators must be propagated.

Together with the partition-function recurrence, these equations 
determine $K(Q)$ using one scalar numerator and $n_\sigma$ auxiliary numerators, 
rather than the quadratic number of individual pair moments. 
The resulting pair-energy average, combined with the mean subshell occupations, 
completes the evaluation of the superconfigurational energy correction 
in Eq.~\eqref{eq:thermal-boltzmann-correction-review}. 
\section{Numerical results}
\label{sec:results}
In this section we present numerical results for two different iron plasmas,
both with a mass density of 0.01~g/cm$^{3}$, and either $kT=$100~eV (case I) or 10~eV (case II).
These are deliberately chosen as regimes where the conventional
alternating recurrences can be used successfully by selecting the appropriate
electron or hole recurrence relations and where neither is reliable through the 
full occupation range. We focus on a single supershell,
$\sigma = \left[3s_{1/2}\cdots 4f_{7/2}\right]$ with a total degeneracy of 50.
Single and two-particle energies as well as UTA shifts and widths are 
computed from relativistic atomic wave functions generated using an average atom 
model implemented in Opus \cite{AbeFenFoo2020}. For the STA moments, we focus in 
particular on the $3d_{5/2}\to4f_{7/2}$ transition. All calculations, unless 
otherwise specified, are performed using double precision (64-bit) arithmetic.

In all figures, solid gray curves show results from direct enumeration.
Orange pluses and green crosses denote the Bar--Shalom recurrences
initialized from empty and full supershells, respectively. Blue circles
denote the electron-counting coefficient passes developed here.
Results from the hole-counting passes are not shown, as they agree with
the electron-counting passes across the entire occupation range in both
cases. We refer to the direct-enumeration results as ``exact.'' To
quantify the accuracy of the conventional recurrences, we define their
working ranges as the contiguous occupation intervals over which 
the relative deviation from direct enumeration does not exceed 1\%. 
Endpoints with vanishing reference values are included when the 
corresponding zeros are reproduced to numerical precision.

\begin{figure}[h]
  \centering
  \includegraphics[width=\myscale\textwidth]{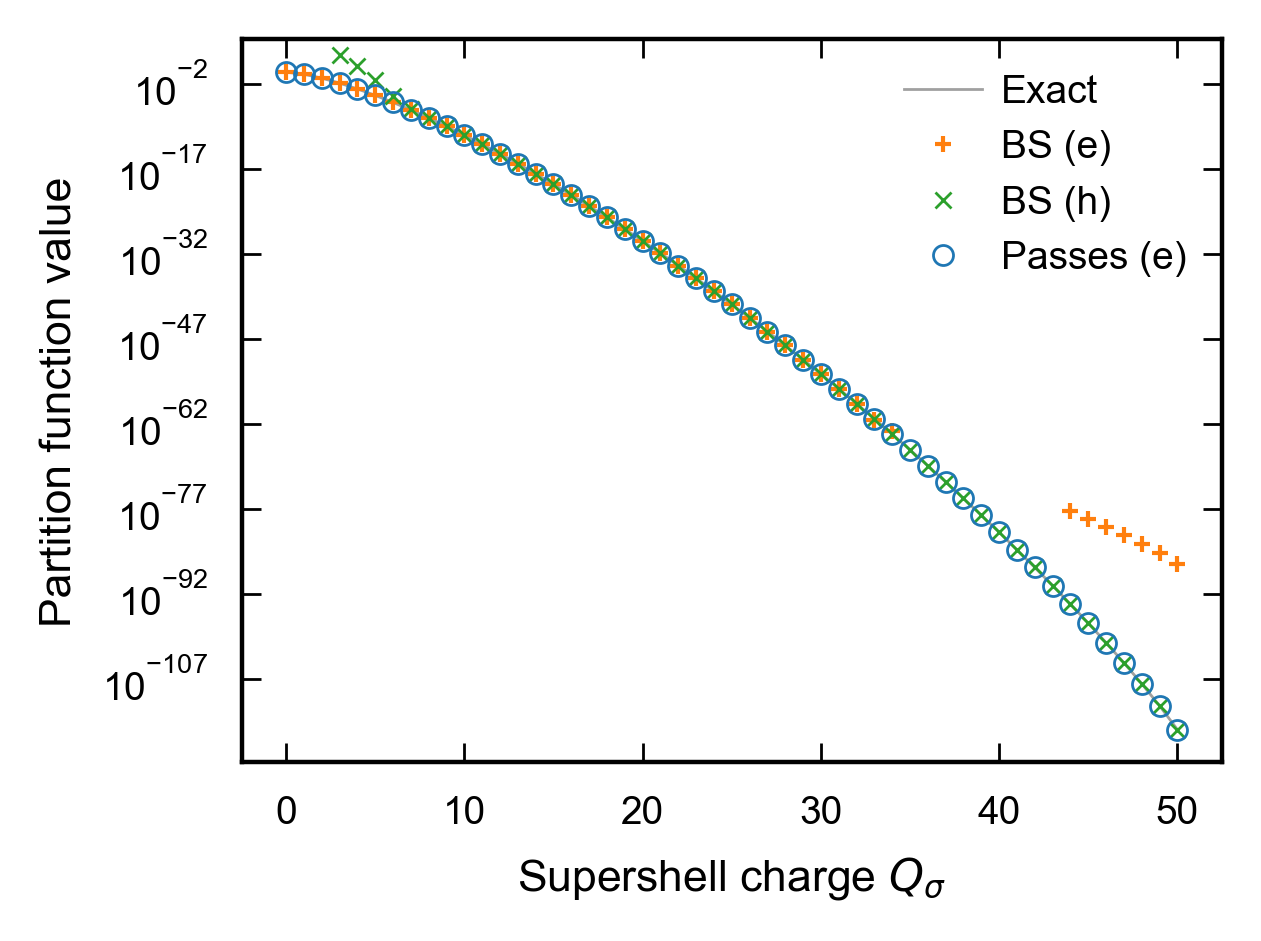} \\
  \includegraphics[width=\myscale\textwidth]{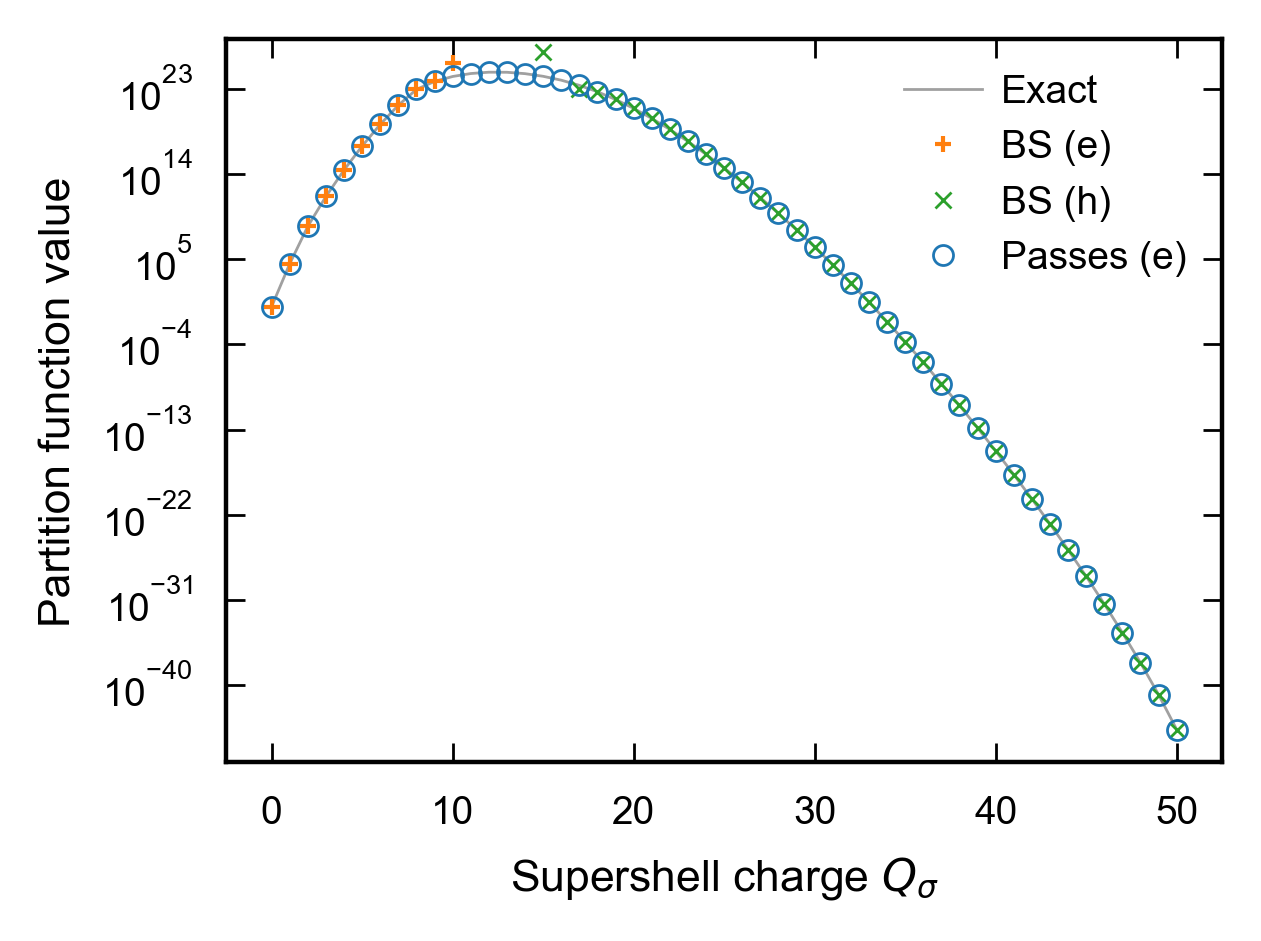}
  \caption{(Color online) Supershell partition functions for the
  $\left[3s_{1/2}\cdots 4f_{7/2}\right]$ supershell. Top panel: Case I, $kT=100$~eV; bottom 
  panel: Case II, $kT=10$~eV. Solid line denotes exact results, orange plus and green cross
  markers indicate Bar--Shalom recurrences from empty and full supershell, respectively,
  and blue circles denote results from the stable  electron recurrences derived in this work.}
  \label{fig:partition-results}
\end{figure}

We first consider the supershell partition function $U_\sigma(Q_\sigma)$.
The top panel of Fig.~\ref{fig:partition-results}, for the 100~eV case,
shows that although the electron- and hole-counting alternating
recurrences of Bar--Shalom and coworkers are not individually accurate
across the entire occupation range, they provide complete coverage
when combined. The electron-counting recurrence agrees with direct
enumeration within 1\% for $0\leq Q_\sigma\leq31$, while the
hole-counting recurrence satisfies the same criterion for
$9\leq Q_\sigma\leq50$. The coefficient-pass recurrence developed here
agrees with direct enumeration throughout the occupation range.

The bottom panel shows $U_\sigma(Q_\sigma)$ for the lower-temperature
case, $kT=10$~eV. The coefficient passes again reproduce the enumerated
results, whereas the working ranges of the alternating recurrences
are restricted to $0\leq Q_\sigma\leq9$ and
$19\leq Q_\sigma\leq50$ for electron and hole counting, respectively.
The recurrence of Gilleron and Pain yields results of comparable
accuracy to the coefficient passes.

We now turn to the supershell contribution to the transition moments.
For Case I, $kT=100$~eV, the Bar--Shalom recurrences for the mean
transition energy and variance exhibit behavior similar to the
partition function: neither electron nor hole counting is accurate
across the entire occupation range, but together they provide
complete coverage. The electron-counting coefficient passes developed
here, on the other hand, reproduce the enumerated moments throughout
the occupation range.

\begin{figure}[h]
  \centering
  \includegraphics[width=\myscale\textwidth]{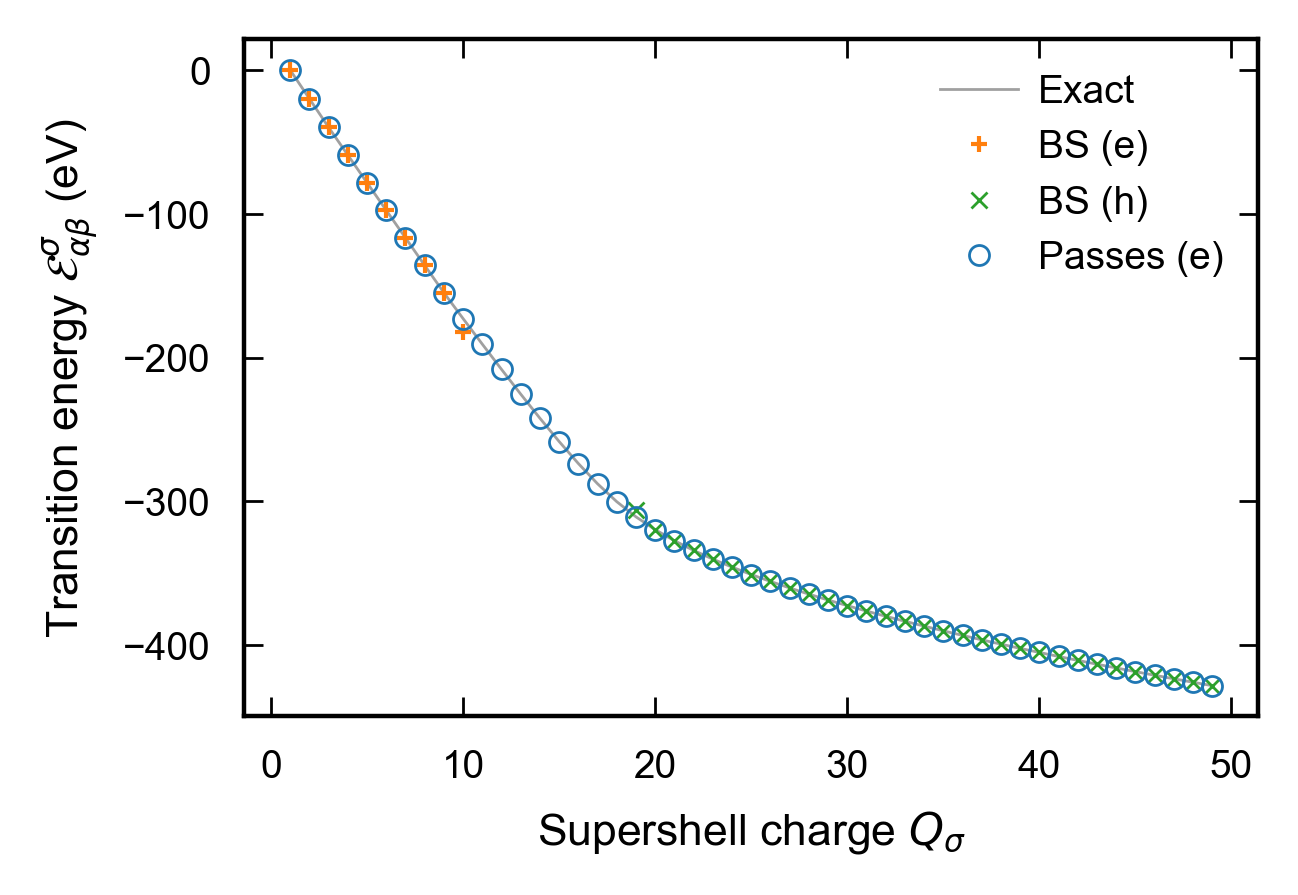}
  \includegraphics[width=\myscale\textwidth]{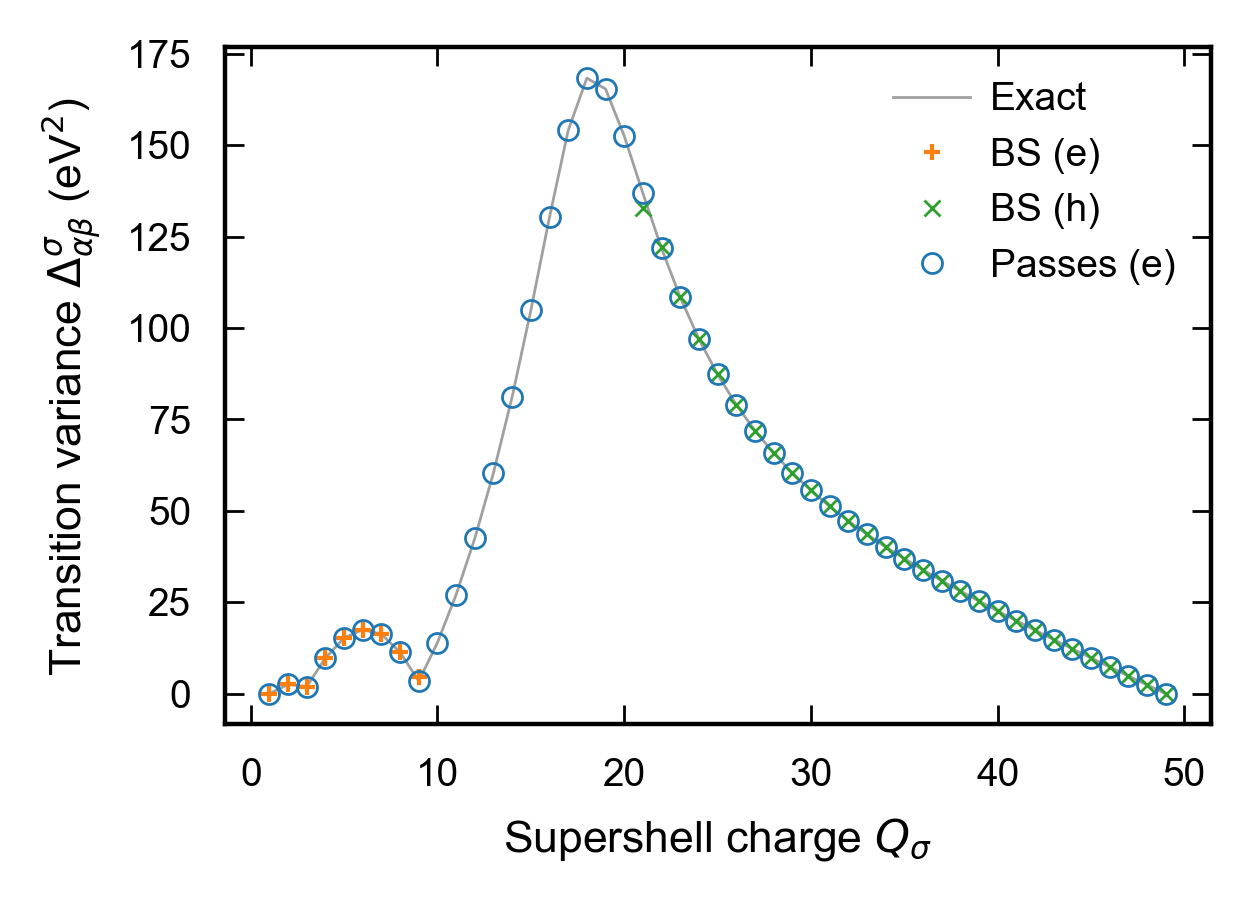}
  \caption{(Color online) Contribution from the $\left[3s_{1/2}\cdots 4f_{7/2}\right]$ supershell
  to the STA mean transition energy (top panel) and variance (bottom panel) of 
  the $3d_{5/2}\to4f_{7/2}$ transition in Case II. The UTA shift and width contributions are included, 
  and the curve and marker conventions follow Fig.~\ref{fig:partition-results}.}
  \label{fig:mean-results}
\end{figure}

The top panel of Fig.~\ref{fig:mean-results} displays the supershell
contribution to the mean transition energy for Case II. Here, the
alternating recurrences have working ranges of
$2\leq Q_\sigma\leq9$ and $20\leq Q_\sigma\leq49$ for electron
and hole counting, respectively. To determine whether the loss of
accuracy originates solely from errors in the partition functions,
we repeated the Bar--Shalom calculations using accurate partition
functions obtained with the Gilleron--Pain method. The results were
essentially unchanged, demonstrating that accurate partition
functions alone are insufficient to prevent cancellation in the
moment calculations.

The bottom panel shows the supershell contribution to the
transition variance, including the internal UTA-width contribution.
The working ranges of the Bar--Shalom recurrences are somewhat narrower:
$2\leq Q_\sigma\leq8$ for electron counting and
$22\leq Q_\sigma\leq49$ for hole counting. In contrast, the
coefficient passes reproduce the enumerated mean energies and
variances throughout the occupation range where these quantities
are defined. We also verified the internal UTA-width contribution 
separately by comparing variances calculated with and without internal 
broadening, keeping the UTA centroid shifts unchanged. The coefficient 
passes reproduce the independently enumerated internal UTA-width 
contributions throughout the tested occupation range. The internal UTA-width 
contribution varies substantially with occupation, accounting for 
approximately 19\% of the total variance at \(Q_\sigma=20\) in Case I and 
1\% in Case II. This agreement independently validates the contracted width 
recurrence, rather than relying solely on agreement of the combined STA variance.

\begin{figure}[h]
  \centering
  \includegraphics[width=\myscale\textwidth]{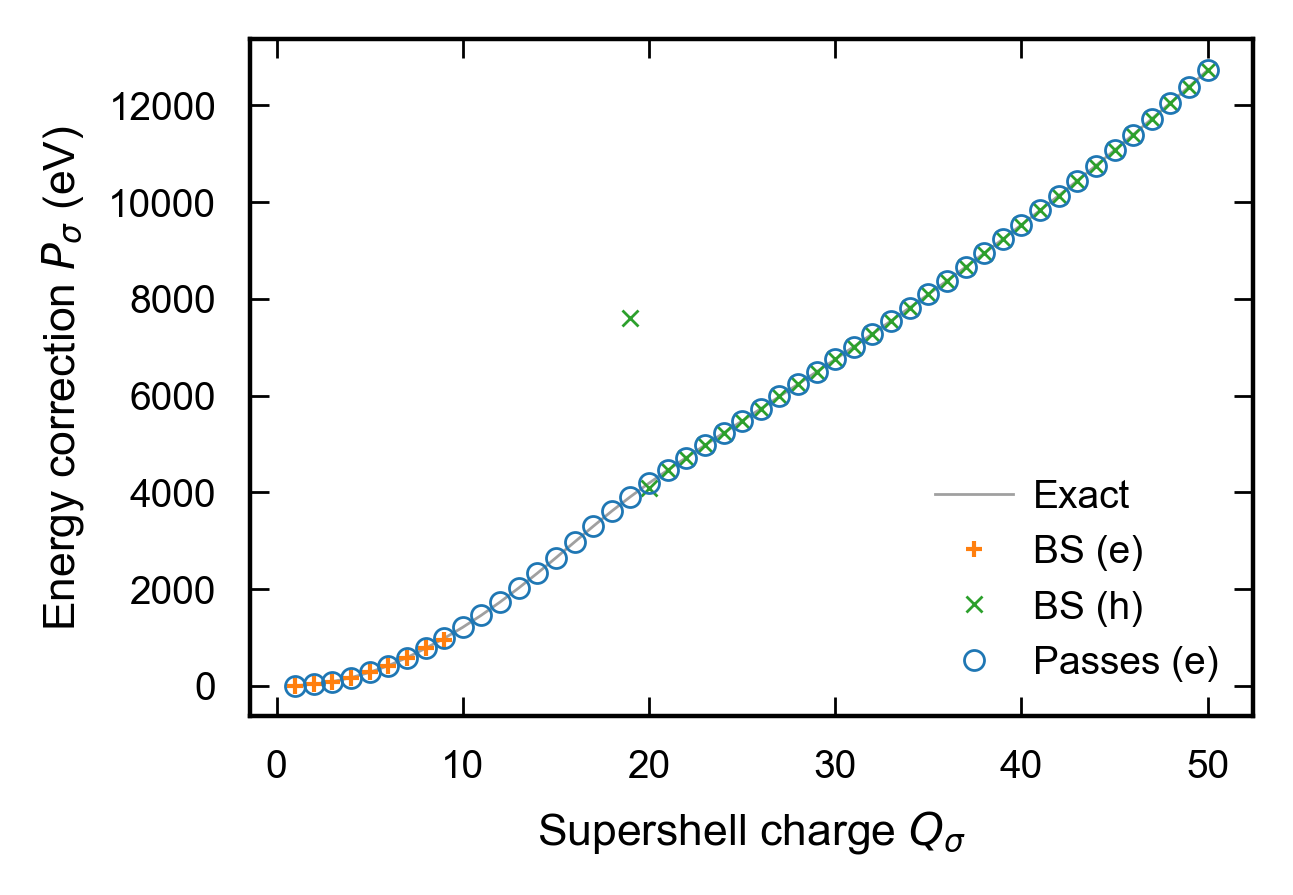}
  \caption{Intra-supershell pair-energy contribution from the $\left[3s_{1/2}\cdots 4f_{7/2}\right]$ 
  supershell in Case II ($kT=10$~eV). The curve and marker conventions follow Fig.~\ref{fig:partition-results}.}
  \label{fig:thermal-results}
\end{figure}

Figure~\ref{fig:thermal-results} shows the intra-supershell
pair-energy contribution entering the Boltzmann energy correction.
For Case II, the working ranges of the Bar--Shalom recurrences are
$2\leq Q_\sigma\leq8$ and $21\leq Q_\sigma\leq50$
for electron and hole counting, respectively. Thus, neither
recurrence provides accurate results in the intermediate occupation
range, while the coefficient passes agree with direct enumeration
throughout.

\begin{figure}[h]
  \centering
  \includegraphics[width=\myscale\textwidth]{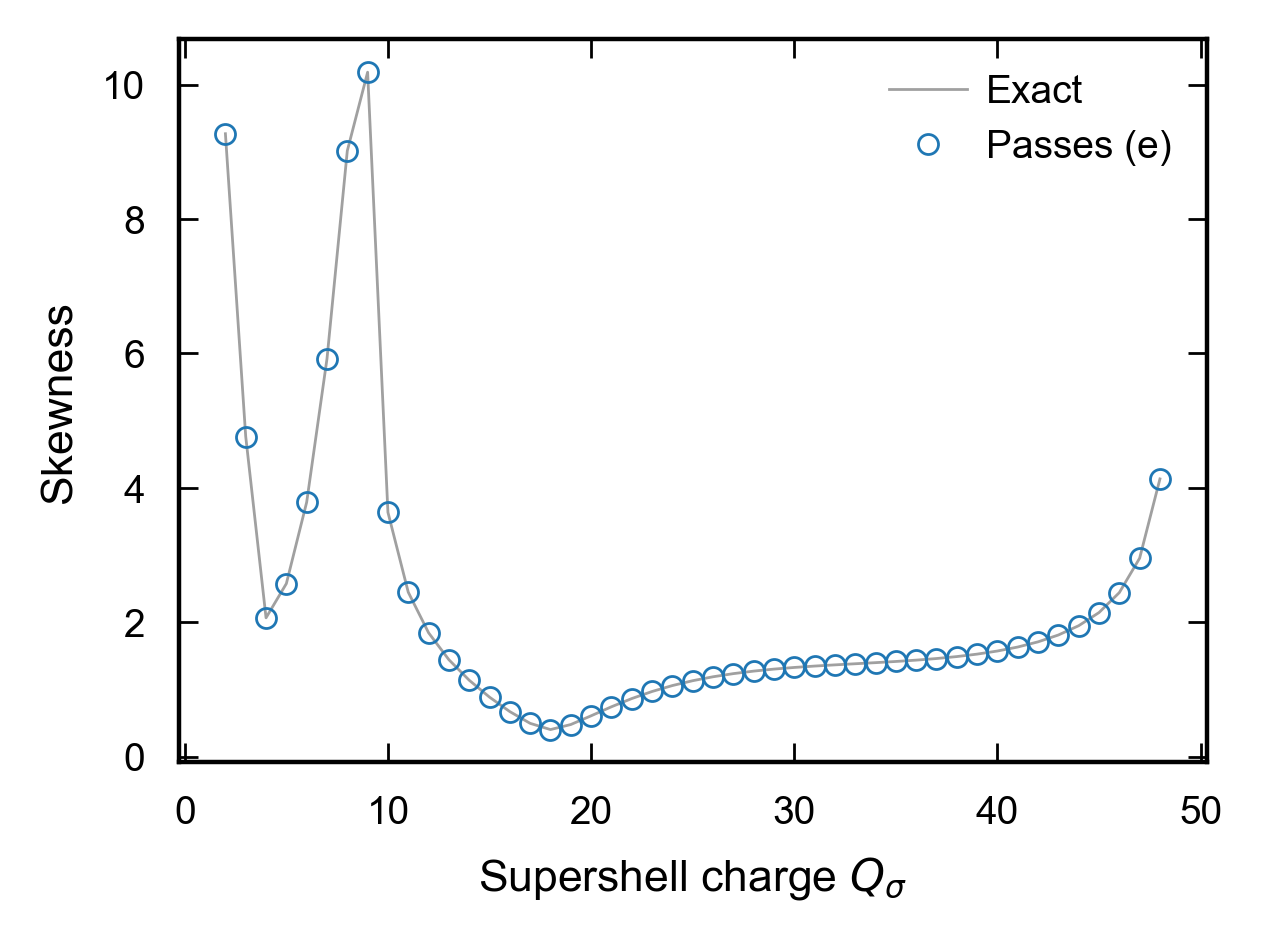}
  \includegraphics[width=\myscale\textwidth]{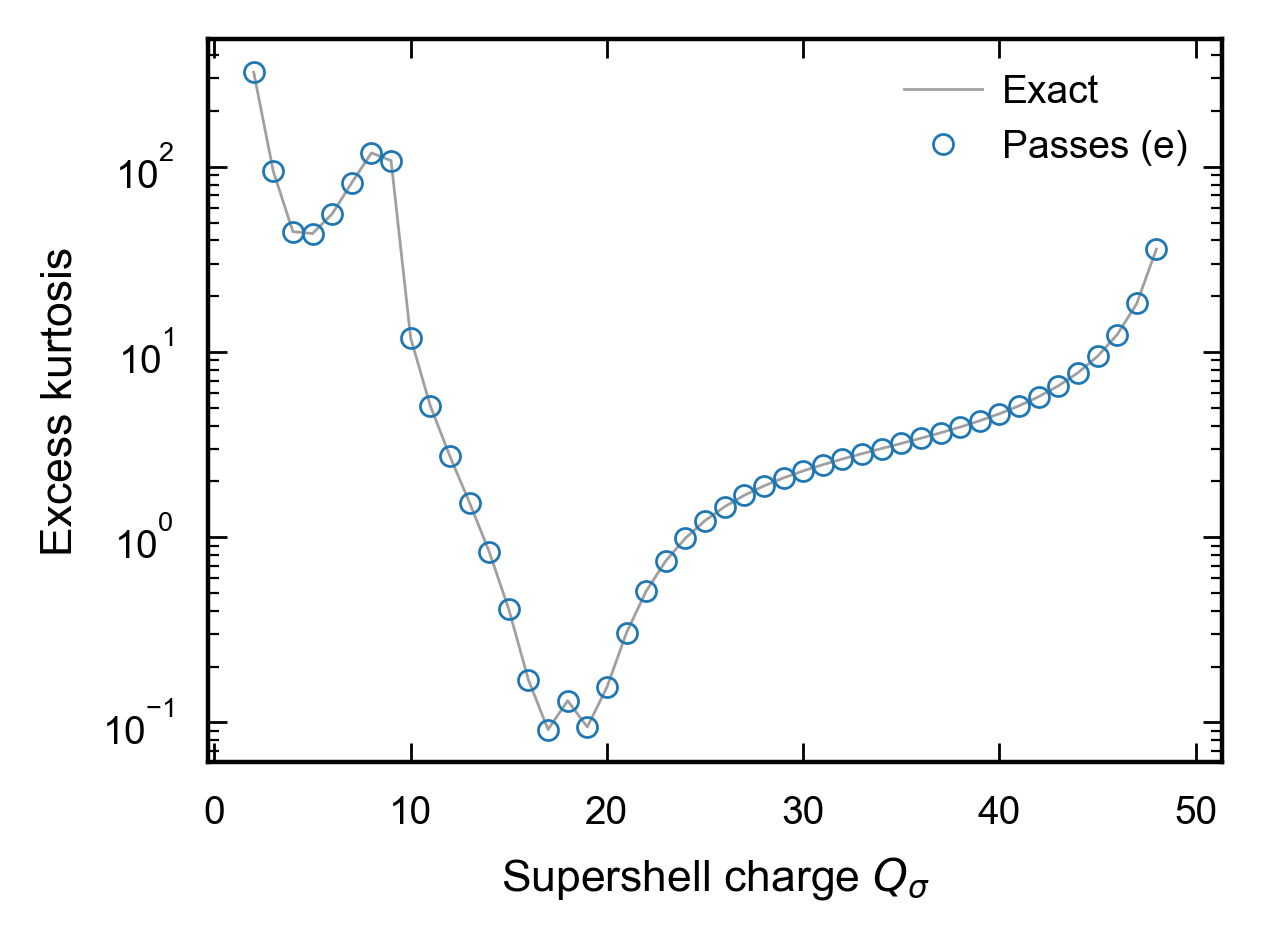}
  \caption{(Color online) Contribution from the $\left[3s_{1/2}\cdots 4f_{7/2}\right]$ supershell
  to the skewness (top panel) and excess kurtosis (bottom panel) of 
  the $3d_{5/2}\to4f_{7/2}$ transition in Case II. The UTA shift and width contributions are 
  not included. Solid line denotes exact results and blue circles denote results from 
  the stable electron recurrences derived in this work.}
  \label{fig:higher-order-results}
\end{figure}

Finally, to demonstrate the applicability of the coefficient passes to higher-order STA moments, 
we compare in Fig.~\ref{fig:higher-order-results} the supershell contribution to the 
skewness and excess kurtosis with values obtained by direct enumeration. The skewness,
\begin{align*}
\gamma_1 = \frac{\mu_3}{\mu_2^{3/2}},
\end{align*}
is the standardized third central moment and measures the asymmetry of the 
transition-energy distribution. The excess kurtosis,
\begin{align*}
\gamma_2 = \frac{\mu_4}{\mu_2^2} - 3,
\end{align*}
is the standardized fourth central moment shifted by three, and measures the departure of 
the distribution's peakedness and tail weight from those of a Gaussian distribution. Thus, 
$\gamma_2=0$ for a Gaussian distribution. Here, we have assumed no other supershell than 
$\left[3s_{1/2}\cdots 4f_{7/2}\right]$ is occupied. The passes reproduce the reference 
results throughout the occupation range where these normalized moments are defined, 
illustrating that the method extends beyond the mean and variance.

These comparisons show that the coefficient-pass recurrences
accurately reproduce the directly enumerated partition functions,
STA energy moments, and configurational-energy corrections throughout
the accessible occupation range. Unlike the conventional alternating
recurrences, the coefficient passes do not require switching between
electron and hole counting to maintain accuracy in the cases studied
here using double precision.

\section{Discussion}
\label{sec:discussion}

The numerical results demonstrate that the coefficient passes provide a stable
alternative to the conventional alternating recurrences for calculating STA
moments and occupation-dependent corrections. In particular, the loss of accuracy
in the Bar--Shalom moment calculations persists even when accurate partition
functions are supplied, indicating that stable evaluation of the partition
function alone is insufficient. The direct coefficient passes, in contrast,
reproduce the enumerated results throughout the tested occupation ranges using
double precision.

Although the coefficient passes avoid the cancellation associated with alternating
reduced-partition expressions, evaluating central moments from raw moments
can still involve subtraction of nearly equal quantities, particularly for
narrow energy distributions or at higher moment orders. 

A central feature of the present formulation is that the different 
statistical averages required for STA calculations can be evaluated 
within the same coefficient-pass framework. Rather than first evaluating 
individual occupation-product averages and subsequently combining them into 
physical observables, the marked generating functions allow the relevant 
combinations to be incorporated directly into closed coefficient recurrences.

Pain and Wilson~\cite{PaiWil2024} derived arbitrary-order centroid-energy 
moments through explicit expansions in occupation products using multinomial 
coefficients and Stirling numbers. The coupled coefficient passes developed 
here replace these explicit expansions by a simple, closed update that 
applies at every moment order, Eq. (49). Moments through order $m$ 
therefore require only $m+1$ coefficient arrays, without enumerating 
mixed occupation products or constructing the corresponding families of 
further reduced-degeneracy partition functions.

A similar reduction in auxiliary quantities is obtained for the 
occupation-dependent corrections. The internal UTA width, expressed in 
Ref.~\cite{PaiWil2024} as a sum of contributions involving further 
reduced-degeneracy partition functions, is accumulated here using the 
partition function, one contracted width numerator, and one 
temporary auxiliary associated with the current subshell. 
The intra-supershell pair-energy contribution requires one contracted 
numerator and $N$ auxiliary numerators, rather than the $N(N+1)/2$ 
individual ordered-pair numerators. Straightforward implementations 
of the respective recurrences require $O(GQ_{\max}m^2)$, $O(GQ_{\max})$, 
and $O(GQ_{\max}N)$ operations. The corresponding working storage 
scales as $O(Q_{\max}m)$, $O(Q_{\max})$, and $O(Q_{\max}N)$.

The arbitrary-order recurrence applies to centroid-energy moments,
as illustrated by the skewness and excess-kurtosis comparisons.
Internal UTA broadening is included here through the second-order
width contribution, rather than through higher-order moments of the
broadened transition distribution. Although the numerical tests are
restricted to individual supershell contributions, the factorization
of the reduced supershell measure allows the centroid-energy cumulants
to be combined across supershells to obtain the corresponding STA moments.
\section{Conclusion}
\label{sec:conclusion} 
We have presented a unified generating-function approach for the 
calculation of super-transition-array moments and occupation-dependent 
averages. By constructing fixed-occupation coefficient passes, we 
obtain coupled recurrences that directly propagate centroid-energy 
moments of arbitrary order without separately evaluating mixed occupation 
products. The same construction provides contracted recurrences for 
UTA-width contributions and superconfigurational energy corrections, 
reducing the number of auxiliary quantities required. Complementary 
electron- and hole-counting formulations allow these quantities to be 
evaluated from either an empty or a completely filled supershell.

Numerical comparisons with direct configuration enumeration demonstrate 
the accuracy of the proposed recurrences for supershell partition 
functions, mean transition energies, variances, higher-order 
centroid-energy moments, and configurational-energy corrections. 
Together, these results provide a common computational framework 
for evaluating the principal statistical averages entering the STA 
formalism, without reverting to separate reduced-partition or 
occupation-product calculations for the different contributions.

\section*{Acknowledgements} 
This work was performed under the auspices of the U.S. Department of Energy 
by Lawrence Livermore National Laboratory under Contract DE-AC52-07NA27344.

\appendix
\section{All-order factorization of reduced STA moments}
\label{app:moment-factorization}

In Sec.~\ref{sec:STA_pass} we derived expressions for the unnormalized 
numerators of the supershell raw STA moments $A_{m,\sigma}$. In this appendix, 
we show how to assemble the mean transition energy and central moments of 
the complete STA.

We begin by recalling that for a fixed superconfiguration 
$\Xi$ and transition $\alpha\to\beta$, the normalized transition moments are

\begin{align}
  \mu_n'
  =
  \frac{
    \displaystyle
    \sum_{C\in\Xi}
    \big( E_{\alpha\beta}^{CC'} \big)^n
    \prod_r 
    \binom{g_r^{\alpha\beta}}{q_r^\alpha}
    X_r^{\,q_r^\alpha}
  }{
    \displaystyle
    \prod_{\sigma\in\Xi}
    U_{Q_\sigma^\alpha}(\mathbf{g}_\sigma^{\alpha\beta})
  }
  \label{eq:preppedMoments}
\end{align}
with the transition energy defined by
\begin{align}
  E_{CC'}^{\alpha\beta}
  &=
  D_0^{\alpha\beta}
  +
  \sum_{\sigma\in\Xi} E_{\alpha\beta}^\sigma,
  \\
  E_{\alpha\beta}^\sigma
  &=
  \sum_{s\in\sigma}
  q_s^\alpha D_s^{\alpha\beta}.
\end{align}

For the mean transition energy, these contributions simply add, while the mixed terms in the 
second raw moment cancel when forming the variance~\cite{BarOreGol1989}. 
At higher orders, however, combining the supershell moments directly becomes increasingly 
complicated.

We will instead show that the cumulants of the complete STA can be expressed as sums of the 
corresponding supershell cumulants, with an additional constant contribution to the first 
cumulant. These cumulants can be constructed from the normalized raw supershell moments 
$A_{m,\sigma}/A_{0,\sigma}$ and subsequently used to obtain the moments of the complete STA.

To that end, we introduce the total superconfigurational moment generating 
function $\mathfrak{F}(t)$ by 
\begin{align}
\mathfrak{F}(t) = \langle e^{t E^{\alpha\beta}_{CC'}} \rangle_\Xi.
\end{align}
As with the marked generating functions introduced in Sec.~\ref{sec:STA_pass}, the 
$n$th raw STA moment can then be obtained by 
\begin{align}
    \mu'_n = \left. \frac{\partial^n \mathfrak{F}(t)}{\partial t^n} \right|_{t=0}.
\end{align}
If $\mathfrak{F}(t)$ is known, we can readily obtain the cumulant generating function
$\mathfrak{K}(t) = \ln\mathfrak{F}(t)$ from which arbitrary cumulants can be obtained 
by
\begin{align}
    \kappa_n = \left. \frac{\partial^n \mathfrak{K}(t)}{\partial t^n} \right|_{t=0}.
\end{align}
To that end, let us form 
\begin{align}
f(t) &=\sum_{n=0}^\infty \frac{t^n}{n!} \mu'_n,
\end{align}
insert Eq.~(\ref{eq:preppedMoments}), change order of summation, and use the
Taylor expansion of $e^x$ to give
\begin{align}
f(t) &= \frac{
    \displaystyle
    \sum_{n=0}^\infty \frac{t^n}{n!} 
    \displaystyle
    \sum_{C\in\Xi}
    \big( E_{\alpha\beta}^{CC'} \big)^n
    \prod_r
    \binom{g_r^{\alpha\beta}}{q_r^\alpha}
    X_r^{\,q_r^\alpha}
  }{
    \displaystyle
    \prod_{\sigma\in\Xi}
    U_{Q_\sigma^\alpha}(\mathbf{g}_\sigma^{\alpha\beta})
  } \nonumber \\
  &=
  \frac{
    \displaystyle
    \sum_{C\in\Xi}
    e^{t E_{\alpha\beta}^{CC'}}
    \displaystyle
    \prod_r
    \binom{g_r^{\alpha\beta}}{q_r^\alpha}
    X_r^{\,q_r^\alpha}
  }{
    \displaystyle
    \prod_{\sigma\in\Xi}
    U_{Q_\sigma^\alpha}(\mathbf{g}_\sigma^{\alpha\beta})
  }
  \label{eq:moment_middle_step}
\end{align}
We recognize this to be exactly the moment generating function $\mathfrak{F}$, and thus 
$f(t)=\mathfrak{F}(t)$.
Before we begin to factorize this expression, we will 
for the sake of brevity use reduced-occupation notation introduced in Sec.~\ref{sec:STA_pass}.
Hence, we will write $q_r=q_r^\alpha$, $g_r=g_r^{\alpha\beta}$, and $Q_\sigma=Q_\sigma^\alpha$.
We will also define $\mathbf q_\sigma=\{q_r\}_{r\in\sigma}$, 
with $|\mathbf q_\sigma|=\sum_{r\in\sigma}q_r$. The notation $\sum_{|\mathbf q_\sigma|=Q_\sigma}$ then 
denotes a sum over all allowed subshell occupations within supershell $\sigma$, subject to $0\leq q_r\leq g_r$.

Since the statistical weights factorize over subshells and the transition energy is a sum 
of supershell contributions, the numerator $p(t)$ of $f(t)$ can be rewritten as  
\begin{align}
    p(t) &= \sum_{C\in\Xi}  e^{tD^{\alpha\beta}_{0}} \prod_{\sigma\in\Xi} \prod_{r\in\sigma} e^{t q_r D^{\alpha\beta}_r}\binom{g_r}{q_r}X_r^{q_r} \nonumber \\
    &= e^{tD^{\alpha\beta}_{0}} \prod_\sigma \Biggl[
         \sum_{|\mathbf{q}_\sigma|=Q_\sigma} 
    \prod_{r\in\sigma} e^{t q_r D^{\alpha\beta}_r}\binom{g_r}{q_r}X_r^{q_r}\Biggr].
\end{align}
As a result, we obtain
\begin{align}
\mathfrak{F}(t) = e^{tD^{\alpha\beta}_{0}} \prod_{\sigma \in \Xi} \mathfrak{F}_\sigma(t),
\label{eq:fullMGFfactorized}
\end{align}
with the normalized supershell moment-generating functions
\begin{align}
\mathfrak{F}_\sigma(t) &= \frac{1}{U_{Q_\sigma}(\mathbf{g}_\sigma)} \nonumber \\
&\times \sum_{|\mathbf{q}_\sigma|=Q_\sigma} 
    \prod_{r\in\sigma} e^{t q_r D^{\alpha\beta}_r}\binom{g_r}{q_r}X_r^{q_r}.
\end{align}
The numerator of this quantity is simply
\begin{align}
[z^{Q_\sigma}]\prod_{r\in\sigma} (1+ze^{tD^{\alpha\beta}_r}X_r)^{g_r} =  [z^{Q_\sigma}]\mathcal{F}_\sigma(z,t),
\end{align}
where $\mathcal{F}_\sigma$ is the unnormalized supershell moment-generating function of Eq.~(\ref{eq:marked-supershell-generating-function}) 
in Sec.~\ref{sec:STA_pass}. We therefore have 
\begin{align}
\mathfrak{F}_\sigma(t) = \frac{[z^{Q_\sigma}]\mathcal{F}_\sigma(z,t)}{A_{0,\sigma}}.
\end{align}

By taking the natural logarithm in Eq.~(\ref{eq:fullMGFfactorized}) we obtain the total superconfigurational cumulant generating function
\begin{align}
\mathfrak{K}(t) &= tD^{\alpha\beta}_{0} +  \sum_{\sigma \in \Xi} \ln \left[ \mathfrak{F}_\sigma(t) \right] \\
&= tD^{\alpha\beta}_{0} +  \sum_{\sigma \in \Xi} \mathfrak{K}_\sigma(t).
\end{align}
As a result, the cumulant-generating function is additive over supershells.
The supershell cumulants are then 
\begin{align}
\kappa_{n} &= \delta_{n1} D^{\alpha\beta}_{0} +  \sum_{\sigma \in \Xi} \kappa_{n,\sigma}.
\label{eq:full_cumulant_from_supershells}
\end{align}

In principle, the supershell cumulants could be obtained by differentiating 
$\mathfrak K_\sigma(t)$ at \(t=0\). Instead, we construct them directly from the 
normalized raw supershell moments already calculated in Sec.~\(\ref{sec:STA_pass}\) 
by using the well-known relation
\begin{align}
    \kappa_{n,\sigma} = r_{n,\sigma} - \sum_{i=1}^{n-1} \binom{n-1}{i-1} \kappa_{i,\sigma} r_{n-i,\sigma},
\end{align}
with $\kappa_{1,\sigma} = r_{1,\sigma}$ and 
\begin{align}
r_{n,\sigma} = \frac{A_{n,\sigma}}{A_{0,\sigma}}.
\end{align}

By Eq.~(\ref{eq:full_cumulant_from_supershells}), the total $n$th cumulant $\kappa_n$ can be 
obtained by summing over supershell cumulants $\kappa_{n,\sigma}$. Finally, the $n$th 
total central moments follow from the recurrence relation 
\begin{align}
\mu_n = \sum_{i=2}^n \binom{n-1}{i-1}\kappa_i \mu_{n-i},
\end{align}
with $\mu_0=1$ and $\mu_1=0$.

As a first few examples, we get the superconfigurational cumulants
\begin{align}
\kappa_{1} &= D_0^{\alpha\beta} + \sum_{\sigma\in\Xi} r_{1,\sigma} \\
\kappa_{2} &= \sum_{\sigma\in\Xi} \left(r_{2,\sigma} - r_{1,\sigma}^2\right)  \\
\kappa_{3} &= \sum_{\sigma\in\Xi} \left(r_{3,\sigma} - 3r_{1,\sigma} r_{2,\sigma} 
   + 2r_{1,\sigma}^3 \right) \\
\kappa_{4}
&= \sum_{\sigma\in\Xi} \left( r_{4,\sigma} - 4r_{1,\sigma}  r_{3,\sigma} 
   + 12r_{1,\sigma}^2r_{2,\sigma} \right. \\
&\quad \left. - 3r_{2,\sigma}^2 - 6r_{1,\sigma}^4 \right).
\end{align}
Therefore, the mean energy is simply given by $\kappa_1$, while $\mu_2=\kappa_2$, 
$\mu_3=\kappa_3$ and $\mu_4=\kappa_4 + 3\kappa_2^2$. 

Thus, the fact that cumulants are additive in supershells provides a 
simple, systematic construction of higher-order STA central moments 
from raw supershell moments.

\section{Hole formalism for STA moments}
\label{app:electron-hole-passes}

In this appendix, we derive complementary hole-counting recurrences
for STA moments, which are useful when a supershell is closer to full
than empty. We continue to use reduced occupations and degeneracies,
and introduce the hole occupations $h_s=g_s-q_s$, with
$G=\sum_s g_s=Q+H$. Let $w=1/z$, $Y_s=1/X_s$, and
\begin{align}
\Lambda &= \prod_s X_s^{g_s},
&
S &= \sum_s g_sD_s.
\end{align}
The marked generating function for STA moments can then be written as
\begin{align}
\mathcal F_\sigma(z,t)
&= \prod_s \left(1+zX_se^{tD_s}\right)^{g_s}
\nonumber\\
&= z^G\Lambda e^{tS}\mathcal F_{\sigma,h}(w,t),
\label{eq:markedF2}
\end{align}
where
\begin{align}
\mathcal F_{\sigma,h}(w,t)
&= \prod_s \left(1+wY_se^{-tD_s}\right)^{g_s}.
\end{align}
Extraction of the electron coefficient at occupation $Q$ therefore
corresponds to the hole coefficient at $H$:
\begin{align}
[z^Q]\mathcal F_\sigma(z,t)
&= \Lambda e^{tS}[w^H]\mathcal F_{\sigma,h}(w,t).
\end{align}

The hole generating function has the same form as the original
electron generating function. As a result, the electron moment
recurrences can be reused with $X_s\to Y_s$, $Q\to H$, and
$D_s\to-D_s$. After $n$ single-state folds, let $C_H^{(n)}(t)$
denote the coefficient of $w^H$ in the partial hole product.
Folding in the next state of subshell $a$ gives
\begin{align}
C_H^{(n+1)}(t)
&= C_H^{(n)}(t)
+Y_ae^{-tD_a}C_{H-1}^{(n)}(t).
\end{align}
By defining
\begin{align}
\widetilde A_m^{(n)}(H)
&=
\left.
\frac{\partial^m C_H^{(n)}(t)}{\partial t^m}
\right|_{t=0},
\end{align}
we obtain the hole recurrence relation for the STA moments:
\begin{align}
\widetilde A_m^{(n+1)}&(H)
=
\widetilde A_m^{(n)}(H)
\nonumber\\
&+
Y_a\sum_{j=0}^{m}\binom{m}{j}
(-D_a)^{m-j}\widetilde A_j^{(n)}(H-1).
\end{align}

We use a tilde to denote moments of the signed hole-energy
contribution itself, while a hat denotes the corresponding electron
moment numerator evaluated through hole counting, with the common
factor $\Lambda$ omitted. Thus,
$A_m(Q)=\Lambda\widehat A_m(H)$.
Applying the product rule to the coefficient identity above gives
\begin{align}
\widehat A_m(H)
&=
\sum_{k=0}^{m}\binom{m}{k}
S^{m-k}\widetilde A_k(H).
\end{align}
In particular, this gives for the first three moments
\begin{align}
\widehat A_0 &= \widetilde A_0,\\
\widehat A_1 &= S\widetilde A_0+\widetilde A_1,\\
\widehat A_2 &= S^2\widetilde A_0
+2S\widetilde A_1+\widetilde A_2.
\end{align}

Alternatively, we can obtain a recurrence relation for
$\widehat A_m$ directly by starting from
\begin{align}
\widehat A_m^{(n)}(H)
&=
\left.
\frac{\partial^m}{\partial t^m}
\left[e^{tS}C_H^{(n)}(t)\right]
\right|_{t=0}.
\end{align}
Multiplying the coefficient recurrence by $e^{tS}$ and applying
the same differentiation gives
\begin{align}
\widehat A_m^{(n+1)}&(H)
=
\widehat A_m^{(n)}(H)
\nonumber\\
&+
Y_a\sum_{j=0}^{m}\binom{m}{j}
(-D_a)^{m-j}\widehat A_j^{(n)}(H-1).
\end{align}
The two recurrence relations for $\widetilde A_m$ and
$\widehat A_m$ are identical, but differ in their initial
conditions. Since $C_H^{(0)}(t)=\delta_{H0}$, we have
\begin{align}
\widetilde A_m^{(0)}(H)
&= \delta_{H0}\delta_{m0},\\
\widehat A_m^{(0)}(H)
&= \delta_{H0}S^m.
\end{align}

It is also convenient to identify
\begin{align}
U_h^{(n)}(H)
&\equiv C_H^{(n)}(0)
=\widetilde A_0^{(n)}(H)
=\widehat A_0^{(n)}(H).
\end{align}
The original electron partition function and normalized moments
are therefore
\begin{align}
U(Q)&=\Lambda U_h(H),\\
\frac{A_m(Q)}{U(Q)}
&=\frac{\widehat A_m(H)}{U_h(H)}.
\end{align}

\section{Hole formalism for subshell occupation moments}
\label{app:electron-hole-passes_occ}

The same hole formalism derived in the previous appendix can also be
used to derive hole recurrences for averages of $f(\mathbf q)$.
Using $h_s=g_s-q_s$ and the definitions of $G$, $H$, $w$, $Y_s$,
and $\Lambda$ introduced there, the marked generating function of
Eq.~(\ref{eq:occ-poly-generator}) can be written as
\begin{align}
\mathcal G_\sigma(z,\mathbf t)
&=
z^G\Lambda e^{\sum_s g_st_s}
\mathcal G_{\sigma,h}(w,-\mathbf t),
\end{align}
where
\begin{align}
\mathcal G_{\sigma,h}(w,\mathbf u)
&=
\prod_s\left(1+wY_se^{u_s}\right)^{g_s}.
\end{align}
The unnormalized numerator is then
\begin{align}
N_f(Q)
&=
[z^Q]\left.
f(\boldsymbol{\partial}_{\mathbf t})
\mathcal G_\sigma(z,\mathbf t)
\right|_{\mathbf t=\mathbf 0}
\nonumber\\
&=
\Lambda[w^H]\left.
f(\mathbf g-\boldsymbol{\partial}_{\mathbf u})
\mathcal G_{\sigma,h}(w,\mathbf u)
\right|_{\mathbf u=\mathbf 0}.
\end{align}
This follows from the fact that $\partial_{t_r}$ acting on
$e^{g_rt_r}A(t_r)$ returns
$e^{g_rt_r}(g_r+\partial_{t_r})A(t_r)$, followed by the
substitution $t_s=-u_s$.

The variables $u_s$ mark the hole occupations in exactly the same
way that the variables $t_s$ mark the electron occupations in the
original generating function. As a result, the recurrence relations
derived for occupation moments can also be used for hole occupations,
replacing $X_s$ by $Y_s=X_s^{-1}$ and $Q$ by $H=G-Q$.
To calculate averages of the original electron occupations, however,
we must account for the transformation $q_s=g_s-h_s$, represented by
$f(\mathbf g-\boldsymbol{\partial}_{\mathbf u})$.

As in the preceding appendix, we define the corresponding electron
numerator evaluated through hole counting, with the common factor
$\Lambda$ removed, as
\begin{align}
\widehat N_f(H)&\equiv \frac{N_f(Q)}{\Lambda}.
\end{align}
Since $U(Q)=\Lambda U_h(H)$, the normalized average is
\begin{align}
\langle f(\mathbf q)\rangle_Q
&=\frac{\widehat N_f(H)}{U_h(H)}.
\end{align}

Suppose that after $n$ single-state folds, the partial hole product
has coefficients $C_H^{(n)}(\mathbf u)$. Folding in the next state
of subshell $a$ gives
\begin{align}
C_H^{(n+1)}(\mathbf u)
&=
C_H^{(n)}(\mathbf u)
+Y_ae^{u_a}C_{H-1}^{(n)}(\mathbf u),
\end{align}
with $U_h^{(n)}(H)=C_H^{(n)}(\mathbf 0)$.
We now illustrate this construction for the occupation passes
already discussed in the main part of this manuscript.

As a first example, consider $f(\mathbf q)=q_r^k$. We define
\begin{align}
\widetilde O_{m,r}^{(n)}(H)
&=
\left.
\frac{\partial^m C_H^{(n)}(\mathbf u)}{\partial u_r^m}
\right|_{\mathbf u=\mathbf 0},
\end{align}
with $\widetilde O_{0,r}^{(n)}(H)=U_h^{(n)}(H)$.
From Eq.~(\ref{eq:occupation-pass}), we can immediately write
\begin{align}
\widetilde O_{m,r}^{(n+1)}&(H)
=
\widetilde O_{m,r}^{(n)}(H)
+Y_a\widetilde O_{m,r}^{(n)}(H-1)
\nonumber\\
&+
\delta_{ar}Y_a
\sum_{p=0}^{m-1}\binom{m}{p}
\widetilde O_{p,r}^{(n)}(H-1).
\label{eq:tildeO_recurrence}
\end{align}
For $m=0$, the sum is empty and this reduces to the
partition-function recurrence.

In the hole representation, the binomial expansion of
$(g_r-\partial_{u_r})^k$ gives
\begin{align}
\widehat O_{k,r}^{(n)}(H)
&=
\sum_{m=0}^{k}(-1)^m\binom{k}{m}
g_r^{k-m}\widetilde O_{m,r}^{(n)}(H).
\end{align}
Rather than assembling the hole moments after completing the
recurrence, we can derive a recurrence relation for the electron
numerator directly. By straightforward manipulations after using
Eq.~(\ref{eq:tildeO_recurrence}), we arrive at
\begin{align}
\widehat O_{k,r}^{(n+1)}&(H)
=
\widehat O_{k,r}^{(n)}(H)
+Y_a\Biggl[
\widehat O_{k,r}^{(n)}(H-1)
\nonumber\\
&+
\delta_{ar}\sum_{j=1}^{k}(-1)^j\binom{k}{j}
\widehat O_{k-j,r}^{(n)}(H-1)
\Biggr].
\end{align}
Contrary to $\widetilde O_{k,r}$, which are numerators of moments
of the hole occupation $h_r$, the quantities $\widehat O_{k,r}$
are numerators of the original electron moments $q_r^k$,
evaluated by a hole pass. Consequently, their initial conditions are
\begin{align}
\widetilde O_{k,r}^{(0)}(H)
&=\delta_{H0}\delta_{k0},\\
\widehat O_{k,r}^{(0)}(H)
&=\delta_{H0}g_r^k,
\end{align}
with $\widehat O_{0,r}^{(n)}(H)=U_h^{(n)}(H)$.
The zero-hole initial conditions for the hatted quantities therefore
correspond to a completely filled supershell.

As a second, much shorter example, consider
$f(\mathbf q)=q_s(g_s-q_s)$. Under the electron-hole transformation,
\begin{align}
q_s(g_s-q_s)&=(g_s-h_s)h_s=h_s(g_s-h_s),
\end{align}
so the polynomial is invariant under exchange of electrons and holes.
Therefore, the previously derived recurrence for the electron-hole
product, Eq.~(\ref{eq:w-recurrence}), can be used directly for holes,
with $X_s\to Y_s$ and $Q\to H$. Here,
$\widehat W_s(H)=\widetilde W_s(H)$ and
$N_f(Q)=\Lambda\widetilde W_s(H)$, with
\begin{align}
\widetilde W_s^{(n+1)}&(H)
=
\widetilde W_s^{(n)}(H)
+Y_a\Bigl[
\widetilde W_s^{(n)}(H-1)
\nonumber\\
&+
\delta_{as}\bigl(
(g_s-1)U_h^{(n)}(H-1)
\nonumber\\
&\qquad
-2\widetilde O_{1,s}^{(n)}(H-1)
\bigr)
\Bigr].
\label{eq:wtilde_rec}
\end{align}
The initial values are $\widetilde W_s^{(0)}(H)=0$.

The third and last example pertains to
$f(\mathbf q)=q_r(q_s-\delta_{rs})$.
The transformed operator can be rearranged into terms we already
know how to handle:
\begin{align}
f(\mathbf g-\boldsymbol{\partial}_{\mathbf u})
&=
g_r(g_s-\delta_{rs})
-(g_s-2\delta_{rs})\partial_{u_r}
\nonumber\\
&\quad
-g_r\partial_{u_s}
+\partial_{u_r}(\partial_{u_s}-\delta_{rs}).
\end{align}
Defining the ordered hole-pair numerator
\begin{align}
\widetilde M_{rs}^{(n)}(H)
&=
\left.
\partial_{u_r}(\partial_{u_s}-\delta_{rs})
C_H^{(n)}(\mathbf u)
\right|_{\mathbf u=\mathbf 0},
\end{align}
we can combine these contributions by linearity to give
\begin{align}
\widehat M_{rs}(H)
&=
g_r(g_s-\delta_{rs})U_h(H)
-g_r\widetilde O_{1,s}(H)
\nonumber\\
&\quad
-(g_s-2\delta_{rs})\widetilde O_{1,r}(H)
+\widetilde M_{rs}(H).
\end{align}
The same relation holds at each stage of the pass.

The ordered hole-pair numerator $\widetilde M_{rs}$ obeys the same
recurrence as its electron-counting counterpart, with $X_a$
replaced by $Y_a$:
\begin{align}
\widetilde M_{rs}^{(n+1)}&(H)
=
\widetilde M_{rs}^{(n)}(H)
+Y_a\Bigl[
\widetilde M_{rs}^{(n)}(H-1)
\nonumber\\
&+
\delta_{ar}\widetilde O_{1,s}^{(n)}(H-1)
+\delta_{as}\widetilde O_{1,r}^{(n)}(H-1)
\Bigr],
\label{eq:tildeM_rec}
\end{align}
with $\widetilde M_{rs}^{(0)}(H)=0$.
Substituting the recurrences for $U_h$, $\widetilde O_{1,r}$,
and $\widetilde M_{rs}$ into the transformation above, and using
$\widehat O_{1,s}=g_sU_h-\widetilde O_{1,s}$, gives
\begin{align}
\widehat M_{rs}^{(n+1)}&(H)
=
\widehat M_{rs}^{(n)}(H)
+Y_a\Bigl[
\widehat M_{rs}^{(n)}(H-1)
\nonumber\\
&-
\delta_{ar}\widehat O_{1,s}^{(n)}(H-1)
-\delta_{as}\widehat O_{1,r}^{(n)}(H-1)
\nonumber\\
&+
2\delta_{ar}\delta_{as}U_h^{(n)}(H-1)
\Bigr].
\label{eq:widehatMrec}
\end{align}
Since the hole pass starts from a completely filled electron
supershell, the initial conditions are
\begin{align}
\widehat M_{rs}^{(0)}(H)
&=\delta_{H0}g_r(g_s-\delta_{rs}).
\end{align}

\subsection{Contracted hole recurrences}

The preceding examples provide hole-counting recurrences for
individual occupation moments and products. As in the electron-counting
formulation, the UTA-width and superconfigurational energy corrections
involve weighted sums of these quantities. The corresponding recurrences
can be contracted, avoiding the explicit evaluation of all individual
contributions. The UTA-width pass uses reduced occupations and
degeneracies, whereas the pair-energy pass uses unreduced ones.

We begin with the UTA-width numerator
\begin{align}
\widetilde L(H)&=\sum_s\Delta_s^2\widetilde W_s(H).
\end{align}
Using Eq.~(\ref{eq:wtilde_rec}), we obtain
\begin{align}
\widetilde L^{(n+1)}(H)
&=
\widetilde L^{(n)}(H)
+Y_a\Bigl[
\widetilde L^{(n)}(H-1)
\nonumber\\
&\qquad
+\Delta_a^2\widetilde V_a^{(n)}(H-1)
\Bigr],
\end{align}
where
\begin{align}
\widetilde V_b^{(n)}(H)
&=
(g_b-1)U_h^{(n)}(H)
-2\widetilde O_{1,b}^{(n)}(H).
\end{align}
The initial condition is $\widetilde L^{(0)}(H)=0$.
As in the electron-counting formulation, only the auxiliary numerator
associated with the current subshell is needed when its spin-orbitals
are folded consecutively. At the beginning of subshell $b$,
$\widetilde O_{1,b}(H)=0$, and hence
\begin{align}
\widetilde V_b(H)&=(g_b-1)U_h(H).
\end{align}
During the subsequent $g_b$ folds, the recurrence simplifies to
\begin{align}
\widetilde V_b^{(n+1)}&(H)
=
\widetilde V_b^{(n)}(H)
\nonumber\\
&+
Y_b\left[
\widetilde V_b^{(n)}(H-1)-2U_h^{(n)}(H-1)
\right].
\end{align}
Once all spin-orbitals of subshell $b$ have been incorporated,
its auxiliary numerator $\widetilde V_b$ is no longer needed.
The calculation therefore requires only $U_h(H)$,
$\widetilde L(H)$, and one temporary auxiliary numerator,
reinitialized at the beginning of each subshell.

Finally, we consider the contracted hole pass of the intra-supershell
pairwise Boltzmann configurational energy,
\begin{align}
\widehat K_\sigma(H_\sigma)
&=
\frac12\sum_{r,s\in\sigma}
\langle r,s\rangle\widehat M_{rs}(H_\sigma).
\end{align}
We again define an auxiliary numerator
\begin{align}
\widehat J_a^{(n)}(H)
&=
\sum_u\langle a,u\rangle\widehat O_{1,u}^{(n)}(H).
\end{align}
Using Eq.~(\ref{eq:widehatMrec}) and suppressing the supershell
label gives
\begin{align}
\widehat K^{(n+1)}(H)
&=
\widehat K^{(n)}(H)
+Y_a\Bigl[
\widehat K^{(n)}(H-1)
\nonumber\\
&\quad
+\langle a,a\rangle U_h^{(n)}(H-1)
-\widehat J_a^{(n)}(H-1)
\Bigr].
\end{align}
The recurrence relation for $\widehat J_t$ is
\begin{align}
\widehat J_t^{(n+1)}(H)
&=
\widehat J_t^{(n)}(H)
+Y_a\Bigl[
\widehat J_t^{(n)}(H-1)
\nonumber\\
&\quad
-\langle t,a\rangle U_h^{(n)}(H-1)
\Bigr].
\end{align}
Since we are starting the recurrence from a full supershell,
the initial conditions are
\begin{align}
\widehat K^{(0)}(H)
&=
\frac{\delta_{H0}}{2}
\sum_{r,s\in\sigma}
\langle r,s\rangle g_r(g_s-\delta_{rs}),\\
\widehat J_t^{(0)}(H)
&=
\delta_{H0}\sum_u\langle t,u\rangle g_u.
\end{align}
These recurrences complete the hole-counting formulation developed
in this work.

\section{Reference unstable recurrences for the superconfigurational energy correction}
\label{app:unstable-boltzmann}
In this Appendix, we supply the unstable recurrence relations 
for the intra-supershell pair-energy, defined in Eq.~(\ref{eq:sc-energy-pair-average}),
as derived by Bar-Shalom~\cite{BarShalom2003Notes}.


For a supershell $ \sigma $ of fixed occupation $ Q_\sigma $, a recurrence 
relation in number of electrons for the local intra-supershell contribution 
may be written as 
\begin{align}
  P_\sigma
  =
  \frac{
    \displaystyle
    \sum_{n=2}^{Q_\sigma}
    U_{Q_\sigma-n}(\mathbf{g}_\sigma)
    F_{n,\mathrm{emp}}^\sigma
  }{
    U_{Q_\sigma}(\mathbf{g}_\sigma)
  }.
  \label{eq:app-boltzmann-empty}
\end{align}
The auxiliary coefficients are
\begin{align}
  F_{n,\mathrm{emp}}^\sigma
  &=
  \frac12\sum_{m=1}^{n-1}
  \sum_{r,s\in\sigma}
  g_r
  \left( g_s - \delta_{rs} \right)
  \langle r,s \rangle \nonumber \\
  &\quad \times
  \left( -X_s \right)^m
  \left( -X_r \right)^{n-m}.
  \label{eq:app-boltzmann-empty-F}
\end{align}

The corresponding recurrence relation in number of holes is 
\begin{align}
  P_\sigma
  =
  \frac{
    \displaystyle
    \sum_{n=0}^{H}U_{Q_\sigma+n}(\mathbf g_\sigma)
    F_{n,\mathrm{full}}^\sigma
  }{U_{Q_\sigma}(\mathbf g_\sigma)},
  \label{eq:app-boltzmann-full}
\end{align}
where $G_\sigma = \sum_{r\in\sigma} g_r$, $H_\sigma = G_\sigma-Q_\sigma$, and 
\begin{align}
  F_{n,\mathrm{full}}^\sigma
  &=\frac12\sum_{m=0}^{n}
  \sum_{r,s\in\sigma}
  \frac{
    g_r\left(g_s-\delta_{rs}\right)\langle r,s\rangle
  }{
    \left(-X_s\right)^m\left(-X_r\right)^{n-m}
  }.
  \label{eq:app-boltzmann-full-F}
\end{align}

%


\begin{thebibliography}{13}%
\makeatletter
\providecommand \@ifxundefined [1]{%
 \@ifx{#1\undefined}
}%
\providecommand \@ifnum [1]{%
 \ifnum #1\expandafter \@firstoftwo
 \else \expandafter \@secondoftwo
 \fi
}%
\providecommand \@ifx [1]{%
 \ifx #1\expandafter \@firstoftwo
 \else \expandafter \@secondoftwo
 \fi
}%
\providecommand \natexlab [1]{#1}%
\providecommand \enquote  [1]{``#1''}%
\providecommand \bibnamefont  [1]{#1}%
\providecommand \bibfnamefont [1]{#1}%
\providecommand \citenamefont [1]{#1}%
\providecommand \href@noop [0]{\@secondoftwo}%
\providecommand \href [0]{\begingroup \@sanitize@url \@href}%
\providecommand \@href[1]{\@@startlink{#1}\@@href}%
\providecommand \@@href[1]{\endgroup#1\@@endlink}%
\providecommand \@sanitize@url [0]{\catcode `\\12\catcode `\$12\catcode
  `\&12\catcode `\#12\catcode `\^12\catcode `\_12\catcode `\%12\relax}%
\providecommand \@@startlink[1]{}%
\providecommand \@@endlink[0]{}%
\providecommand \url  [0]{\begingroup\@sanitize@url \@url }%
\providecommand \@url [1]{\endgroup\@href {#1}{\urlprefix }}%
\providecommand \urlprefix  [0]{URL }%
\providecommand \Eprint [0]{\href }%
\providecommand \doibase [0]{https://doi.org/}%
\providecommand \selectlanguage [0]{\@gobble}%
\providecommand \bibinfo  [0]{\@secondoftwo}%
\providecommand \bibfield  [0]{\@secondoftwo}%
\providecommand \translation [1]{[#1]}%
\providecommand \BibitemOpen [0]{}%
\providecommand \bibitemStop [0]{}%
\providecommand \bibitemNoStop [0]{.\EOS\space}%
\providecommand \EOS [0]{\spacefactor3000\relax}%
\providecommand \BibitemShut  [1]{\csname bibitem#1\endcsname}%
\let\auto@bib@innerbib\@empty
\bibitem [{\citenamefont {{Bar-Shalom}}\ \emph {et~al.}(1989)\citenamefont
  {{Bar-Shalom}}, \citenamefont {Oreg}, \citenamefont {Goldstein},
  \citenamefont {Shvarts},\ and\ \citenamefont {Zigler}}]{BarOreGol1989}%
  \BibitemOpen
  \bibfield  {author} {\bibinfo {author} {\bibfnamefont {A.}~\bibnamefont
  {{Bar-Shalom}}}, \bibinfo {author} {\bibfnamefont {J.}~\bibnamefont {Oreg}},
  \bibinfo {author} {\bibfnamefont {W.~H.}\ \bibnamefont {Goldstein}}, \bibinfo
  {author} {\bibfnamefont {D.}~\bibnamefont {Shvarts}},\ and\ \bibinfo {author}
  {\bibfnamefont {A.}~\bibnamefont {Zigler}},\ }\href
  {https://doi.org/10.1103/PhysRevA.40.3183} {\bibfield  {journal} {\bibinfo
  {journal} {Phys. Rev. A}\ }\textbf {\bibinfo {volume} {40}},\ \bibinfo
  {pages} {3183} (\bibinfo {year} {1989})}\BibitemShut {NoStop}%
\bibitem [{\citenamefont {{Bar-Shalom}}\ \emph {et~al.}(1995)\citenamefont
  {{Bar-Shalom}}, \citenamefont {Oreg},\ and\ \citenamefont
  {Goldstein}}]{BarOreGol1995}%
  \BibitemOpen
  \bibfield  {author} {\bibinfo {author} {\bibfnamefont {A.}~\bibnamefont
  {{Bar-Shalom}}}, \bibinfo {author} {\bibfnamefont {J.}~\bibnamefont {Oreg}},\
  and\ \bibinfo {author} {\bibfnamefont {W.~H.}\ \bibnamefont {Goldstein}},\
  }\href {https://doi.org/10.1103/PhysRevE.51.4882} {\bibfield  {journal}
  {\bibinfo  {journal} {Phys. Rev. E}\ }\textbf {\bibinfo {volume} {51}},\
  \bibinfo {pages} {4882} (\bibinfo {year} {1995})}\BibitemShut {NoStop}%
\bibitem [{\citenamefont {{Bauche-Arnoult}}\ \emph {et~al.}(1979)\citenamefont
  {{Bauche-Arnoult}}, \citenamefont {Bauche},\ and\ \citenamefont
  {Klapisch}}]{BauBauKla1979}%
  \BibitemOpen
  \bibfield  {author} {\bibinfo {author} {\bibfnamefont {C.}~\bibnamefont
  {{Bauche-Arnoult}}}, \bibinfo {author} {\bibfnamefont {J.}~\bibnamefont
  {Bauche}},\ and\ \bibinfo {author} {\bibfnamefont {M.}~\bibnamefont
  {Klapisch}},\ }\href {https://doi.org/10.1103/PhysRevA.20.2424} {\bibfield
  {journal} {\bibinfo  {journal} {Phys. Rev. A}\ }\textbf {\bibinfo {volume}
  {20}},\ \bibinfo {pages} {2424} (\bibinfo {year} {1979})}\BibitemShut
  {NoStop}%
\bibitem [{\citenamefont {{Bauche-Arnoult}}\ \emph {et~al.}(1982)\citenamefont
  {{Bauche-Arnoult}}, \citenamefont {Bauche},\ and\ \citenamefont
  {Klapisch}}]{BauBauKla1982}%
  \BibitemOpen
  \bibfield  {author} {\bibinfo {author} {\bibfnamefont {C.}~\bibnamefont
  {{Bauche-Arnoult}}}, \bibinfo {author} {\bibfnamefont {J.}~\bibnamefont
  {Bauche}},\ and\ \bibinfo {author} {\bibfnamefont {M.}~\bibnamefont
  {Klapisch}},\ }\href {https://doi.org/10.1103/PhysRevA.25.2641} {\bibfield
  {journal} {\bibinfo  {journal} {Phys. Rev. A}\ }\textbf {\bibinfo {volume}
  {25}},\ \bibinfo {pages} {2641} (\bibinfo {year} {1982})}\BibitemShut
  {NoStop}%
\bibitem [{\citenamefont {{Bauche-Arnoult}}\ \emph {et~al.}(1985)\citenamefont
  {{Bauche-Arnoult}}, \citenamefont {Bauche},\ and\ \citenamefont
  {Klapisch}}]{BauBauKla1985}%
  \BibitemOpen
  \bibfield  {author} {\bibinfo {author} {\bibfnamefont {C.}~\bibnamefont
  {{Bauche-Arnoult}}}, \bibinfo {author} {\bibfnamefont {J.}~\bibnamefont
  {Bauche}},\ and\ \bibinfo {author} {\bibfnamefont {M.}~\bibnamefont
  {Klapisch}},\ }\href {https://doi.org/10.1103/PhysRevA.31.2248} {\bibfield
  {journal} {\bibinfo  {journal} {Phys. Rev. A}\ }\textbf {\bibinfo {volume}
  {31}},\ \bibinfo {pages} {2248} (\bibinfo {year} {1985})}\BibitemShut
  {NoStop}%
\bibitem [{\citenamefont {Oreg}\ \emph {et~al.}(1997)\citenamefont {Oreg},
  \citenamefont {{Bar-Shalom}},\ and\ \citenamefont
  {Klapisch}}]{OreBarKla1997}%
  \BibitemOpen
  \bibfield  {author} {\bibinfo {author} {\bibfnamefont {J.}~\bibnamefont
  {Oreg}}, \bibinfo {author} {\bibfnamefont {A.}~\bibnamefont {{Bar-Shalom}}},\
  and\ \bibinfo {author} {\bibfnamefont {M.}~\bibnamefont {Klapisch}},\ }\href
  {https://doi.org/10.1103/PhysRevE.55.5874} {\bibfield  {journal} {\bibinfo
  {journal} {Phys. Rev. E}\ }\textbf {\bibinfo {volume} {55}},\ \bibinfo
  {pages} {5874} (\bibinfo {year} {1997})}\BibitemShut {NoStop}%
\bibitem [{\citenamefont {Wilson}\ and\ \citenamefont
  {Chen}(1999)}]{WilChe1999}%
  \BibitemOpen
  \bibfield  {author} {\bibinfo {author} {\bibfnamefont {B.~G.}\ \bibnamefont
  {Wilson}}\ and\ \bibinfo {author} {\bibfnamefont {M.~H.}\ \bibnamefont
  {Chen}},\ }\href {https://doi.org/10.1016/S0022-4073(98)00069-7} {\bibfield
  {journal} {\bibinfo  {journal} {J. Quant. Spectrosc. Radiat. Transf.}\
  }\textbf {\bibinfo {volume} {61}},\ \bibinfo {pages} {813} (\bibinfo {year}
  {1999})}\BibitemShut {NoStop}%
\bibitem [{\citenamefont {Gilleron}\ and\ \citenamefont
  {Pain}(2004)}]{GilPai2004}%
  \BibitemOpen
  \bibfield  {author} {\bibinfo {author} {\bibfnamefont {F.}~\bibnamefont
  {Gilleron}}\ and\ \bibinfo {author} {\bibfnamefont {J.-C.}\ \bibnamefont
  {Pain}},\ }\href {https://doi.org/10.1103/PhysRevE.69.056117} {\bibfield
  {journal} {\bibinfo  {journal} {Phys. Rev. E}\ }\textbf {\bibinfo {volume}
  {69}},\ \bibinfo {pages} {056117} (\bibinfo {year} {2004})}\BibitemShut
  {NoStop}%
\bibitem [{\citenamefont {Wilson}\ \emph {et~al.}(2007)\citenamefont {Wilson},
  \citenamefont {Gilleron},\ and\ \citenamefont {Pain}}]{WilGilPai2007}%
  \BibitemOpen
  \bibfield  {author} {\bibinfo {author} {\bibfnamefont {B.~G.}\ \bibnamefont
  {Wilson}}, \bibinfo {author} {\bibfnamefont {F.}~\bibnamefont {Gilleron}},\
  and\ \bibinfo {author} {\bibfnamefont {J.-C.}\ \bibnamefont {Pain}},\ }\href
  {https://doi.org/10.1103/PhysRevE.76.032103} {\bibfield  {journal} {\bibinfo
  {journal} {Phys. Rev. E}\ }\textbf {\bibinfo {volume} {76}},\ \bibinfo
  {pages} {032103} (\bibinfo {year} {2007})}\BibitemShut {NoStop}%
\bibitem [{\citenamefont {Pain}\ \emph {et~al.}(2020)\citenamefont {Pain},
  \citenamefont {Gilleron},\ and\ \citenamefont {Wilson}}]{PaiGilWil2020}%
  \BibitemOpen
  \bibfield  {author} {\bibinfo {author} {\bibfnamefont {J.-C.}\ \bibnamefont
  {Pain}}, \bibinfo {author} {\bibfnamefont {F.}~\bibnamefont {Gilleron}},\
  and\ \bibinfo {author} {\bibfnamefont {B.~G.}\ \bibnamefont {Wilson}},\
  }\href {https://doi.org/10.1016/j.hedp.2020.100891} {\bibfield  {journal}
  {\bibinfo  {journal} {High Energy Density Phys.}\ }\textbf {\bibinfo {volume}
  {37}},\ \bibinfo {pages} {100891} (\bibinfo {year} {2020})}\BibitemShut
  {NoStop}%
\bibitem [{\citenamefont {Pain}\ and\ \citenamefont
  {Wilson}(2024)}]{PaiWil2024}%
  \BibitemOpen
  \bibfield  {author} {\bibinfo {author} {\bibfnamefont {J.-C.}\ \bibnamefont
  {Pain}}\ and\ \bibinfo {author} {\bibfnamefont {B.~G.}\ \bibnamefont
  {Wilson}},\ }\href {https://doi.org/10.1016/j.hedp.2024.101104} {\bibfield
  {journal} {\bibinfo  {journal} {High Energy Density Phys.}\ }\textbf
  {\bibinfo {volume} {51}},\ \bibinfo {pages} {101104} (\bibinfo {year}
  {2024})}\BibitemShut {NoStop}%
\bibitem [{\citenamefont {{Bar-Shalom}}(2003)}]{BarShalom2003Notes}%
  \BibitemOpen
  \bibfield  {author} {\bibinfo {author} {\bibfnamefont {A.}~\bibnamefont
  {{Bar-Shalom}}}} (\bibinfo {year} {2003}),\ \bibinfo {note} {unpublished
  notes, dated July 13, 2003}\BibitemShut {NoStop}%
\bibitem [{\citenamefont {\AA{}berg}\ \emph {et~al.}(2020)\citenamefont
  {\AA{}berg}, \citenamefont {Fenn}, \citenamefont {Foord}, \citenamefont
  {Grabowski}, \citenamefont {Iglesias},\ and\ \citenamefont
  {Wilson}}]{AbeFenFoo2020}%
  \BibitemOpen
  \bibfield  {author} {\bibinfo {author} {\bibfnamefont {D.}~\bibnamefont
  {\AA{}berg}}, \bibinfo {author} {\bibfnamefont {D.}~\bibnamefont {Fenn}},
  \bibinfo {author} {\bibfnamefont {M.~E.}\ \bibnamefont {Foord}}, \bibinfo
  {author} {\bibfnamefont {P.~E.}\ \bibnamefont {Grabowski}}, \bibinfo {author}
  {\bibfnamefont {C.~A.}\ \bibnamefont {Iglesias}},\ and\ \bibinfo {author}
  {\bibfnamefont {B.~G.}\ \bibnamefont {Wilson}},\ }\href
  {https://doi.org/10.2172/1650424} {\emph {\bibinfo {title} {The {{Opus}}
  Radiative Opacity Code, a New Implementation of the Super-Transition Array
  Method}}},\ \bibinfo {type} {Tech. Rep.}\ \bibinfo {number} {LLNL-TR-813832}\
  (\bibinfo {address} {United States},\ \bibinfo {year} {2020})\BibitemShut
  {NoStop}%
\end{thebibliography}
\end{document}